\documentclass[lettersize,journal]{IEEEtran}
\usepackage[utf8]{inputenc}

\usepackage{xspace}
\usepackage{url}
\usepackage{graphicx}
\usepackage{epstopdf}
\usepackage{paralist}
\usepackage[normalem]{ulem}
\usepackage{fancyref} % \fref

\usepackage{amsmath}
\usepackage{amsfonts}
\usepackage{amssymb}

\usepackage{bm}

\usepackage{textcomp}
\usepackage{array}
\usepackage{arydshln}
\usepackage{mwe}
\usepackage{url}
\usepackage{siunitx} 
\usepackage{tikz}
\usepackage{gensymb}

\usepackage[inline]{enumitem}
\usepackage{xcolor}
\definecolor{links}{rgb}{0.3,0,0}   % red
\definecolor{urls}{rgb}{0,0,0.8}    % blue
\definecolor{cites}{rgb}{0,0,0.6}   % blue
\usepackage[colorlinks,hyperindex,linkcolor=links,citecolor=cites,urlcolor=urls]{hyperref} % generates colored links in pdf file

\definecolor{silver}{cmyk}{0,0,0,0.3}
\definecolor{navy}{cmyk}{0.8,0.5,0,0}
\definecolor{lightblue}{cmyk}{0.35,0.11,0,0}
\definecolor{orange}{cmyk}{0,0.57,0.86,0}
\definecolor{yellow}{cmyk}{0,0,0.9,0.0}
\definecolor{reddishyellow}{cmyk}{0,0.22,1.0,0.0}
\definecolor{lightred}{cmyk}{0,0.820,0.753,0.0}
\definecolor{black}{cmyk}{0,0,0.0,1.0}
\definecolor{white}{cmyk}{0,0,0.0,0}
\definecolor{purple}{cmyk}{0.64,0.83,0,0}
\definecolor{darkYellow}{cmyk}{0,0,1.0,0.5}
\definecolor{darkSilver}{cmyk}{0,0,0,0.1}
\definecolor{lightyellow}{cmyk}{0,0,0.3,0.0}
\definecolor{lighteryellow}{cmyk}{0,0,0.1,0.0}
\definecolor{lightestyellow}{cmyk}{0,0,0.05,0.0}
\definecolor{darkblue}{cmyk}{0.98,0.89,0,0.11}
\definecolor{bluel1}{cmyk}{0.5,0.05,0.05,0.05}
\definecolor{darkred}{cmyk}{0,0.89,0.7,0.55}
\definecolor{magenta}{rgb}{1.0, 0.0, 1.0}
\definecolor{cyan}{rgb}{0.0, 1.0, 1.0}
\usepackage{dsfont}
\usepackage{footnote}
\usepackage{balance}
\usepackage{float}
\usepackage{cite}

\usepackage{wasysym}
\usepackage{amssymb}
\usepackage{tabularx,booktabs}
\newcolumntype{L}[1]{>{\raggedright\let\newline\\\arraybackslash\hspace{0pt}}m{#1}}
\newcolumntype{C}[1]{>{\centering\let\newline\\\arraybackslash\hspace{0pt}}m{#1}}
\newcolumntype{R}[1]{>{\raggedleft\let\newline\\\arraybackslash\hspace{0pt}}m{#1}}
\usepackage{boldline}

\usepackage{algorithmicx}
\usepackage{algorithm}
\usepackage{algpseudocode}
\algnewcommand\ForGiven{\textbf{for a given }}
\algnewcommand\algorithmiccompute{\textbf{Compute }}
\algnewcommand\Compute{\item[\algorithmiccompute]}

\usepackage{subcaption}
\usepackage{pgfplots}
\usepgfplotslibrary{groupplots}
\usetikzlibrary{pgfplots.groupplots}

\usepackage{vmr-symbols-vecbold}
\usepackage{standard-macros}

\newcommand{\nl}{\mathrm{L}}
\newcommand{\nk}{\mathrm{K}}
\newcommand{\rate}{\mathrm{R}}
\newcommand{\mN}{\mathrm{N}}
\newcommand{\mM}{\mathrm{M}}

\safemath{\rmatW}{\mathbb{W}}
\safemath{\rmatY}{\mathbb{Y}}
\safemath{\rmatZ}{\mathbb{Z}}
\safemath{\matR}{\mathsf{R}}
\safemath{\matZ}{\mathsf{Z}}
\safemath{\matQ}{\mathsf{Q}}
\safemath{\matI}{\mathsf{I}}
\safemath{\matY}{\mathsf{Y}}

\let\ns\undefined
\newcommand{\nb}{n_{\text{b}}}
\newcommand{\ns}{n_\text{{s}}}
\newcommand{\nc}{n_{\text{c}} }
\newcommand{\np}{n_{\text{p}} }
\newcommand{\infden}{\imath_{s}}

\safemath{\tR}{\tilde{\rate}}
\safemath{\tgamma}{\tilde{\gamma}}
\safemath{\mfn}{\mathfrak{N}_{n,\gamma''(\zeta)}}
\safemath{\mfnt}{\tilde{\mathfrak{N}}_{n,\gamma''(\zeta)}}
\newcommand{\mNsim}{\mathsf{N}\sub{sim}}
\newcommand{\hatVech}{\skew{-5}\hat{\vech}}

\IEEEoverridecommandlockouts
\begin{document}
\title{Efficient evaluation of the error probability for pilot-assisted URLLC with Massive MIMO}
\author{\IEEEauthorblockN{A.~Oguz Kislal, Alejandro Lancho,~\IEEEmembership{Member,~IEEE}, Giuseppe Durisi,~\IEEEmembership{Senior Member,~IEEE}, and Erik G.~Ström,~\IEEEmembership{Fellow,~IEEE}}
% \IEEEauthorblockA{\IEEEauthorrefmark{1}Department of Electrical Engineering,
% Chalmers University of Technology, 41296 Gothenburg, Sweden\\}
% \IEEEauthorblockA{\IEEEauthorrefmark{2}Department of Electrical Engineering and Computer Science,
% Massachusetts Institute of Technology, Cambridge, MA, USA}

\thanks{A. Oguz Kislal, Giuseppe Durisi, and Erik. G. Str\"om are with the Department of
Electrical Engineering, Chalmers University of Technology, Gothenburg 41296,
Sweden (e-mail: \{kislal,durisi,erik.strom\}@chalmers.se). Alejandro Lancho is with the Department of Electrical Engineering and Computer Science, Massachusetts Institute of Technology, Cambridge 02139, MA, USA (e-mail: lancho@mit.edu). 
This work was partly supported by the Swedish Research Council under Swedish Research Council grant 2018-04359. 
Alejandro Lancho has received funding from the European Union's Horizon 2020 research and innovation programme under the Marie Sklodowska-Curie grant agreement No. 101024432. This work was also supported by the National Science Foundation under Grant No CCF-2131115.
}
\thanks{This work was presented in part at the Asilomar Conf. Signals,
Syst., Comput., Pacific Grove CA, U.S.A., Nov. 2022~\cite{kislal22-11z}.}
}
% \date{}
\maketitle
\sloppy
\begin{abstract}
We propose a numerically efficient method for evaluating the random-coding union bound with parameter $s$ on
the error probability achievable in the finite-blocklength regime by a pilot-assisted transmission scheme
employing Gaussian codebooks and operating over a memoryless block-fading channel. Our method relies on the saddlepoint
approximation, which, differently from previous results reported for similar scenarios, is performed with
respect to the number of fading blocks (a.k.a. diversity branches) spanned by each codeword, instead of the
number of channel uses per block. This different approach avoids a costly numerical averaging of the error
probability over the realizations of the fading process and of its pilot-based estimate at the receiver and
results in a significant reduction of the number of channel realizations required to estimate the error
probability accurately. Our numerical experiments for both single-antenna communication
links and massive multiple-input multiple-output (MIMO) networks show that, when
two or more diversity branches are available, the error probability can be
estimated accurately with the saddlepoint approximation with respect to the
number of fading blocks using a numerical method that requires about two orders
of magnitude fewer Monte-Carlo samples than with the saddlepoint approximation with respect
to the number of channel uses per block.
\end{abstract}
%% ----------------------------------------------------------------------------


\section{Introduction}
Next-generation wireless communication systems will support mission-critical links operating under stringent
reliability and latency constraints.
Denoted as ultra-reliable low-latency communications (URLLC), this type of links will enable applications such as
vehicle-to-everything communication, factory automation~\cite{Ren2020}, autonomous driving~\cite{Song2019},
and haptic communications~\cite{Berg2017}.

One crucial characteristic of the URLLC traffic is that it often involves small information payloads combined
with short packets, i.e., packets consisting of a small number of coded symbols.
To understand why short packets are needed, it is worth recalling that the length of a data packet depends on the product of the available bandwidth and the signal
duration.
In URLLC, the signal duration is limited because of the latency constraint of
the targeted applications (e.g., control of automated factories, critical
internet-of-things services). 
The bandwidth is often also
limited, because of the need to orthogonalize the transmission of different users to avoid multiuser
interference, which has a negative impact on the packet error probability.
As pointed out in, e.g.,~\cite{durisi16-09a}, the classic asymptotic performance metrics used to design communication systems, i.e., the ergodic and the outage rates, are unsuitable in the short-packet regime. Thus, 
a much more precise characterization of the tradeoff between transmission rate and error probability is required.

Finite-blocklength information theory, a field whose relevance to URLLC has become apparent after the seminal
works in~\cite{hayashi09-11a,Polyanskiy2010}, provides a precise characterization of such tradeoffs, in terms of nonasymptotic upper (achievability) and lower (converse) bounds on the smallest error probability compatible with a given SNR, transmission rate and packet size.

To satisfy the reliability requirements over fading channels in URLLC, under the above-mentioned diversity limitations in both time and frequency, 
it becomes crucial to leverage on the spatial diversity offered by multiple
antennas. 
A promising approach is to use massive multiple-input
multiple-output (MIMO)---a wireless cellular network architecture in which a base station (BS) with a large number of active
antennas serves multiple users on the same time-frequency resources~\cite{Marzetta2010}. 
The benefits of massive MIMO are well understood~\cite{EmilBook2017}, and this
technology has been incorporated into the 5G standard. 
%One challenge is that the available literature is mostly based on the assumption of ergodic (e.g. \cite{Bjornson2018}) or quasistatic (e.g. \cite{Karlsson2018}) regimes, and such assumptions are questionable for the URLLC scenarios.  

Focusing on communication over memoryless block-fading channels, we present in this paper a numerically
efficient method to evaluate information-theoretic upper bounds on the
finite-blocklength error probability achievable in practically relevant
scenarios, including massive MIMO deployments. 
Methods such as the one
presented in this paper are necessary
since evaluating most of the available information-theoretic
error-probability bounds and approximations that are accurate
for scenarios of interest for URLLC~\cite{Ostman2019,Lancho2020,ostman2020}
is---as we shall see---extremely time consuming.
This prevents the use of such expressions within  
URLLC optimization routines such as resource-allocation and scheduling
algorithms.

\paragraph*{State of the art}
Throughout the paper, we will focus on the upper bound on the error probability obtained by using the
random-coding union bound with parameter $s$ (RCUs) proposed in~\cite{Martinez2011}.
As discussed in, e.g.,~\cite{Ostman2019}, this bound is particularly suited for transmission over fading
channels because it provides achievability results that hold both for the optimal noncoherent maximum-likelihood (ML) decoder, as well as
for more practically relevant transmission schemes that rely on pilot-assisted transmission (PAT). For example, PAT schemes include the case in which
the acquired channel estimate at the receiver is treated as perfect via the use of a mismatched scaled
nearest-neighbor (SNN) decoder~\cite{Lapidoth2002}. In fact, for this setup, it appears that the RCUs bound is the only known tractable bound on the error probability. Moreover, the normal and saddlepoint approximations seem to be the only reasonable ways to approximate the RCUs bound.

The RCUs bound involves the computation of a certain tail probability, which is not known in closed form and needs to
be evaluated numerically. If performed naively, this step is time consuming because of the low error probabilities of
interest in URLLC.
A common approach to circumvent this issue encompasses the following two steps.
One starts by noting that, given the
realization of the fading channel and of its estimate at the receiver, the random variable whose tail probability is
of interest can be written as a sum of independent random variables. 
Then, one uses the central-limit theorem
to approximate this tail probability by a Gaussian tail probability.
The resulting approximation, which is typically referred to as \emph{normal
approximation}~(see,
e.g.,\cite[Sec.IV]{Polyanskiy2010}), is, however, not accurate for the error probabilities of interest in the URLLC regime~\cite{Martinez2011,ostman2020}.
Furthermore, this approach still requires one to perform a Monte-Carlo averaging over the
channel realizations and their estimate at the receiver, which is time consuming.

As shown in, e.g.,~\cite{Martinez2011,font-segura18-03a}, a much more accurate approximation can be obtained
by using the so-called \textit{saddlepoint method}~\cite{jensen95-a}. 
Consider a memoryless block-fading
channel where each packet is assumed to span $\nb$ fading blocks. Assume that, within each fading block, $\ns$
coded symbols are transmitted. For this scenario, the saddlepoint method can be applied in two different ways:
we can fix $\nb$ and perform a saddlepoint expansion with respect to (w.r.t.)
$\ns$, i.e., perform an expansion that is accurate when
the number of symbols per block is
large.
Alternatively, we can fix $\ns$ and perform a saddlepoint expansion w.r.t $\nb$,
i.e., perform an expansion that is
accurate when the number of fading blocks (a.k.a. diversity branches) is large.

For PAT transmission and SNN decoding, the first approximation has been recently
studied in~\cite{Ostman2019,ostman2020}
for the special case $\nb=1$.
As discussed in~\cite{ostman2020}, this approach yields an approximation on the conditional error probability given
the channel and its estimate, which needs then to be averaged w.r.t. the channel realizations and their estimates
at the receiver.
A different approach to evaluate this conditional error probability is described
in~\cite{Taricco2022}. 

The second approximation was studied in~\cite{Lancho2020}, but only for the case of optimal ML decoder. This approximation pertains the unconditional tail probability, and, hence, does not require an additional averaging over the realizations of the channel and its estimate.
As we shall see, this makes the numerical evaluation of this approximation
computationally efficient. 

\paragraph*{Contributions}
Focusing on independent and identically
distributed (\iid) Gaussian codebooks, we present in this paper two saddlepoint
approximations on the RCUs for the practically relevant case of PAT and SNN decoding: the one w.r.t. $\ns$ generalizes the one
reported in~\cite{ostman2020} to arbitrary $\nb$ values;
the one w.r.t. to $\nb$ generalizes the one reported in~\cite{Lancho2020} to PAT and SNN decoding.
Considering the URLLC regime, we then provide a detailed analysis of the
accuracy and the computational complexity of both approximations in: 
\begin{inparaenum}[i)]
\item a single-input single-output (SISO) setup; 
\item the uplink of a two-user single-cell massive MIMO network; 
\item the uplink of a multi-user
multi-cell massive MIMO network.
\end{inparaenum}
This progression allows us to understand the
impact of the number of BS and users in the network on both
the accuracy and the numerical complexity of the considered approximations.
Our analysis
shows that, despite being developed under the assumption of large $\nb$, the
saddlepoint w.r.t. $\nb$ is accurate for $\nb$ values as small as $2$ in both
SISO and multi-user MIMO scenarios.
Furthermore, it entails a much smaller computational complexity than the saddlepoint w.r.t. $\ns$.
Specifically, whenever $\nb\geq 2$,  the number of samples required to evaluate
the saddlepoint w.r.t. $\nb$ via Monte-Carlo simulation is typically around $2$ orders
of magnitude smaller than the number of samples required to evaluate the
saddlepoint approximation w.r.t. $\ns$, once the averaging over the
channel and its estimate is accounted for.
We also show that, for the scenarios considered in the paper, the normal
approximation is typically not accurate.
%% ----------------------------------------------------------------------------
\paragraph*{Notation} 
We denote random vectors and random scalars by upper-case boldface letters such as
$\rvecx$ and upper-case standard letters,
such as $\rndx$, respectively. 
Their realizations are indicated by lower-case letters of the same font. 
We use upper-case letters of two special fonts to denote deterministic matrices
(e.g., $\matY$) and random matrices (e.g., $\rmatY$). 
To avoid
ambiguities, we use another font, such as $\rate$ for rate, to denote constants that are typically
capitalized in the literature. 
The identity matrix of size $a \times a$ is written as $\matI_a$. The circularly-symmetric Gaussian distribution is denoted by $\jpg(0,
\sigma^2)$, where $\sigma^2$ denotes the variance. The superscripts $(\cdot)^T$,
$(\cdot)^H$, and
$(\cdot)^*$ denote transposition, Hermitian transposition, and complex conjugation, respectively. We write $\log(\cdot)$ 
to denote the natural logarithm, $\vecnorm{\cdot}$ stands for the $\ell^2$-norm, $\Prob[\cdot]$
for the probability of an event, $\Exop[\cdot]$ for the expectation operator, $\Var{\cdot}$ for the variance of a random variable,  
and $Q(\cdot)$ for the Gaussian
$Q$-function. 
Finally, for two functions $f(n)$ and $g(n)$, the notation $f(n) = o(g(n))$ means that $\lim_{n\to\infty}
f(n)/g(n)=0$ and the notation $f(n) =\mathcal{O}(g(n))$ means that $\limsup_{n\to\infty} \abs{f(n)/g(n)}<\infty$.

\paragraph*{Organization of the paper}
In Section~\ref{sec:non_asymp_SISO}, we present a finite-blocklength upper bound on the error
probability for the SISO Rayleigh block-fading
channel. We then introduce different methods to evaluate this bound in the URLLC regime. The extension of our framework to multicell,
multiuser massive MIMO networks is presented in Section \ref{sec:MultiCell}. In Section \ref{Sec:NumericalResults}, we discuss the accuracy and computational complexity of the methods introduced in Section~\ref{sec:non_asymp_SISO} with the help of numerical examples.
Concluding remarks are provided in Section~\ref{sec:conclusion}.
%% ----------------------------------------------------------------------------
\section{A Non-Asymptotic Upper Bound on the Error Probability}
\label{sec:non_asymp_SISO}

\subsection{The SISO System Model}
\label{sec:System}
\begin{figure}[t]
    \centering
    \includegraphics[width=.95\columnwidth]{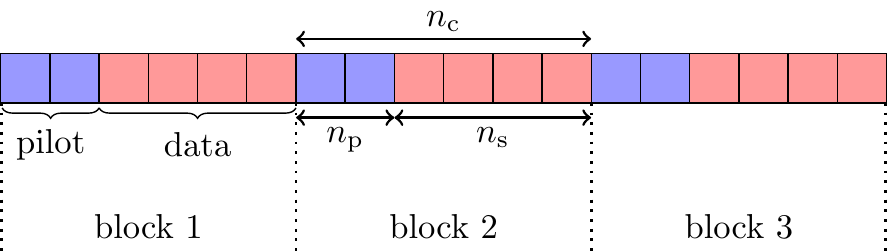}
    \caption{Structure of a packet for $\nc = 6$, $\np = 2$, $\nb = 3$.}
    \label{fig:PacketFig}
\end{figure}

We start by considering a SISO memoryless block-fading channel.
Specifically, the channel is assumed to stay constant over the transmission of a block of $\nc$ channel uses and to change
independently across blocks.
Each transmitted packet spans $\nb$ such fading blocks. 
Hence, each packet consists of $\nb\nc$ complex-valued symbols.
We assume that the first $\np$ symbols within each block are pilots known to the
receiver and the remaining $\ns = \nc - \np$ symbols are data symbols, as illustrated in Fig.~\ref{fig:PacketFig}.
Pilot symbols are used to
estimate the fading channel within the corresponding block.
The input-output relation corresponding to the pilot transmission phase within block $\ell=1,\dots,\nb$ is %can be 
modeled as 
\begin{equation}\label{eq:io-relation-siso}
    \rvecy_{\ell}^{(p)} = \rndh_{\ell}\vecx_{\ell}^{(p)} + \rvecw_{\ell}^{(p)}.
\end{equation}
Here, $\vecx_{\ell}^{{p}}$ denotes the deterministic $\np$-dimensional vector of pilot symbols, which we assume to satisfy the
power constraint $\vecnorm{\vecx_{\ell}^{(p)}}^{2}=\rho\np$, where $\rho$ denotes the average transmit power per symbol. %transmitted power.
Furthermore, $\rndh_{\ell}$ denotes the scalar random fading complex channel gain, and
$\rvecw_{\ell}^{(p)}$  denotes the $\np$-dimensional
additive noise vector, which may depend on the fading process.\footnote{Allowing
for such dependency will turn out crucial to extend the SISO analysis to the
multiuser MIMO case.} 
We assume that the entries
of $\rvecw_{\ell}^{(p)}$ are conditionally independent and
$\jpg(0,\sigma^2_{\ell} )$-distributed given the
realization of the fading process. 

The received vector $\rvecy_{\ell}^{(p)}$ and the pilot sequence $\vecx_{\ell}^{(p)}$ are used by the receiver
to obtain an estimate~$\hat{\rndh}_{\ell}$ of the channel~$\rndh_{\ell}$. 
Note that we have not specified the fading distribution or the algorithm used by
the receiver to estimate the fading channel.
Indeed, the error probability bounds we shall present in this section 
hold for arbitrary fading distributions and arbitrary channel-estimation algorithms.

Within each block, the pilot-transmission phase is followed by a data-transmission phase involving
$\ns=\nc-\np$ symbols per block, i.e., a total of $\nb\ns$ symbols.
The input-output relation for the $\ell$th block in the data phase is given by 
\begin{equation}
    \label{eq:SISOdata}
    \rvecy_{\ell} = \rndh_{\ell}\vecx_{\ell} + \rvecw_{\ell}.
\end{equation}
We assume that the $\nb\ns$-dimensional vector $[\tp{\vecx_{1}},\dots,
\tp{\vecx_{\nb}}]^T$ is selected from a codebook
$\setC$ of size $\ceil{\exp(\nb\nc\rate)}$, where $\rate$ denotes the
transmission rate in nats per channel use.\footnote{With an abuse of
notation, we will use $\rate$ to denote also the rate measured in bits per
channel use  when presenting numerical experiments in
Sections~\ref{sec:SISO-experiments} and~\ref{sec:numerical-experiments-2-users}}
%The noise vector $\rvecw_{\ell}$ may be dependent of the fading process and, given the realization of the fading process, it consist of i.i.d. $\jpg(0,\sigma^2)$ entries. 

To perform decoding, the receiver seeks the codeword in the codebook that is closest to the received
signal, once each part of the codeword corresponding to a different fading block is scaled by the available
channel estimate.
Mathematically, given the received vector $[\tp{\vecy_{1}},\dots,
\tp{\vecy_{\nb}}]^T$ and the channel estimates
$\{\hat{h}_{1},\dots,\hat{h}_{\nb}\}$, the decoded codeword
$\hat{\vecx}=[\tp{\hat{\vecx}_{1}},\dots, \tp{\hat{\vecx}_{\nb}}]^T$ is determined as
follows:
\begin{equation}
    \hat{\vecx} = \argmin_{\bar{\vecx}=\tp{[\tp{\bar{\vecx}_{1}},\dots,\tp{\bar{\vecx}_{\nb}}]}  \in\setC} \sum_{\ell=1}^{\nb} \vecnorm{\vecy_{\ell}
  -\hat{h}_{\ell}\bar{\vecx}_{\ell}}^{2}. \label{eq:snn_dec}
\end{equation}
This decoder, which is known as mismatched SNN decoder, coincides with the ML decoder only when the receiver has perfect
channel-state information, i.e., $\hat{h}_{\ell}=h_{\ell}$
for $\ell=1,\dots,\nb$.
The attractive feature of this decoder is that information-theoretic bounds on
its error probability  
can be approached in practice using good channel codes for the nonfading AWGN channel~\cite{Ostman2019}.
In contrast, approaching information-theoretic error-probability bounds for the optimal ML decoder considered
in~\cite{Lancho2020}
with low-complexity coding schemes is still an open problem (note, however, the recent progress reported in~\cite{yuan21-06l}).

%% ----------------------------------------------------------------------------
\subsection{The RCUs Finite-Blocklength Bound}
Like most of the achievablity results in information theory, the RCUs bound~\cite{Martinez2011} we shall focus on in this paper 
is obtained by means of a random-coding argument. Specifically, one evaluates the average error probability averaged over a randomly constructed ensemble of codebooks.
In this paper, we consider the \iid Gaussian ensemble, in which each symbol of each codeword is drawn independently from a $\jpg(0,\rho)$ distribution, where
$\rho$ models the average transmit power per symbol in the data phase (same power as in the pilot phase). 
Although suboptimal at finite blocklength~\cite{scarlett16-10a} in the nonfading SISO case,  the \iid Gaussian ensemble is often used in the literature because it leads tractable expressions when applied to PAT, SNN decoding, and multiuser MIMO scenarios.

Specialized to our setup, the RCUs bound results in the following upper bound $\epsilon\sub{ub}$ on the \emph{packet error probability} $\epsilon$:
\begin{equation}
    \label{eq:RCUsBound}
    \epsilon \leq \epsilon\sub{ub}=\Prob\ltrsqr{\frac{\log \rndu}{\nc\nb} +  \frac{1}{\nc\nb}
  \sum_{\ell=1}^{\nb}\infden\ltrp{\rvecx_{\ell};\rvecy_{\ell},\hat{\rndh}_{\ell}} \leq
  \rate}.
\end{equation}
Here, $\rndu$ is a random variable that is uniformly distributed on $[0,1]$ and
independent of all other quantities,
\begin{equation}\label{eq:decomposition-is-vec}
   \infden\ltrp{\rvecx_{\ell};\rvecy_{\ell},\hat{\rndh}_{\ell}} =\sum_{k=1}^{\ns}
   \infden\ltrp{\rndx_{k,\ell}; \rndy_{k,\ell}, \hat{\rndh}_{\ell}}    
\end{equation}
where $\rndx_{k,\ell}$ and $\rndy_{k,\ell}$ are the $k$th element of $\rvecx_{\ell}$ and
$\rvecy_{\ell}$ respectively, and 
$\infden(x;y,\hat{h})$
%$\infden\ltrp{\rndx_{k,\ell}; \rndy_{k,\ell}, \hat{\rndh}_{\ell}}$ 
is the so-called \emph{generalized
information density}, which, for the case of \iid $\jpg(0,\rho)$ codebooks and SNN
decoding, is given by~\cite[App.~A]{ostman2020} 
\begin{IEEEeqnarray}{rCl}
    \nonumber
  \infden(x;y,\hat{h}) &=&  
-s \abs{y - \hat{h}x}^2 \\
                       &&+ \frac{s\abs{ y}^2}{1+s\rho\abs{\hat{h}}^2 }+ \log\ltrp{1+ s\rho\abs{ \hat{h}}^2}.
\end{IEEEeqnarray}
Finally, $s>0$ is an optimization parameter that can be used to tighten the bound. 

In general, no closed-form expression is available for~\eqref{eq:RCUsBound}.
Hence, this probability needs to be
evaluated with numerical methods.
A na\"ive implementation of this step results in time-consuming simulations, given the low target error probabilities of
interest in URLLC.
We next discuss two approaches to compute~\eqref{eq:RCUsBound}
efficiently: one is based on
asymptotic expansions of~\eqref{eq:RCUsBound} applied w.r.t. the number of data symbols per block $\ns$,
and the other is based on asymptotic expansions of~\eqref{eq:RCUsBound}
applied w.r.t. the number of blocks $\nb$. 
For each approach, we will present an expansion based on the central-limit
theorem, which will result in the so-called normal approximation, and an expansion based on the saddlepoint method.

%% ----------------------------------------------------------------------------
\subsection{Asymptotic Expansion w.r.t. $\ns$}
\label{sec:ns-expansions}
The idea behind this approach, which for the case $\nb=1$  has been explored in~\cite{ostman2020}, is to
analyze first a conditional version of the probability in~\eqref{eq:RCUsBound}, 
in which the channel and its estimate within each block are given.
Specifically, one focuses on 
\begin{multline}
  \epsilon_{\text{ub}}\ltrp{\vech,\hatVech}= \Prob\biggl[ \frac{\log \rndu}{\nc\nb} +  \frac{1}{\nc\nb}
  \sum_{\ell=1}^{\nb} \sum_{k=1}^{\ns}
   \infden\ltrp{\rndx_{k,\ell}; \rndy_{k,\ell}, \hat{h}_{\ell}}\\  \leq R \big|   \rvech=\vech, \hat{\rvech}=\hatVech
 \biggr]\label{eq:conditional_prob} 
\end{multline}
where $\rvech=[\rndh_{1},\dots, \rndh_{\nb}]^T$ and
$\hat{\rvech}=[\hat{\rndh}_{1},\dots, \hat{\rndh}_{\nb}]^T$.\footnote{Note that $\rndy_{k,\ell}$ depends on $\rndh_{\ell}$ via~\eqref{eq:SISOdata}.}
Then one seeks an asymptotic approximation to this conditional probability. The approximations that will be presented in this section are easy to evaluate numerically because they depend on quantities that can be evaluated in closed form. To obtain the desired estimate of $\epsilon_{\text{ub}}$, one still needs to perform the averaging
\begin{IEEEeqnarray}{rCL}
  \epsilon_{\text{ub}} = \Ex{\rvech,\hat{\rvech}} {\epsilon_{\text{ub}}\ltrp{\rvech,\hat{\rvech}}}\label{eq:RCUs_ns_expect}
\end{IEEEeqnarray}
over the channel and its estimate numerically.
\subsubsection{Normal Approximation w.r.t. $\ns$}\label{sec:normal-ns}
One way to numerically approximate~\eqref{eq:conditional_prob} is to perform a normal approximation w.r.t. $\ns$
based on the Berry-Esseen central-limit theorem [20, Ch. XVI.5].
Specifically, note that given $\rvech$ and $\hat{\rvech}$, the $\nb\ns$ random variables
$\bigl\{\infden(\rndx_{k,\ell}; \rndy_{k,\ell}, \hat{h}_{\ell})\bigr\}$
in~\eqref{eq:conditional_prob} are conditionally independent and identically distributed within each fading block. Let $I_s(h_\ell,\hat{h}_\ell) = \Exop[\infden(\rndx_{1,\ell},\rndy_{1,\ell},\hat{h}_\ell)]$  and $V_s(h_{\ell}, \hat{h}_{\ell}) = \Varop[\infden(\rndx_{1,\ell};\rndy_{1,\ell},\hat{h}_{\ell})]$ denote the mean and the variance of the information density, respectively. Then, conditioned on $\rvech = \vech$ and $\hat{\rvech} = \hat{\vech}$ it follows that
\begin{IEEEeqnarray}{rCl}
    \IEEEeqnarraymulticol{3}{l}{
    \nonumber
    \Ex{}{\sum_{\ell=1}^{\nb}\sum_{k=1}^{\ns} \infden\ltrp{\rndx_{k,\ell};\rndy_{k,\ell},\hat{h}_\ell}} } \nonumber \\ \qquad \qquad \qquad \qquad 
    &=& \ns \sum_{\ell=1}^{\nb}\Ex{}{\infden\ltrp{\rndx_{1,\ell},\rndy_{1,\ell},\hat{h}_\ell}}
    \label{eq:NormalIs_proof_step1}
    \\
    &=& \ns \sum_{\ell=1}^{\nb} I_s\ltrp{h_\ell,\hat{h}_\ell}
    \label{eq:NormalIs_proof}
\end{IEEEeqnarray}
and 
\begin{IEEEeqnarray}{rCl}
    \IEEEeqnarraymulticol{3}{l}{
    \nonumber
    \Var{\sum_{\ell=1}^{\nb}\sum_{k=1}^{\ns} \infden\ltrp{\rndx_{k,\ell};\rndy_{k,\ell},\hat{h}_\ell}} } \nonumber \\ \qquad\qquad\qquad
    &=& \ns \sum_{\ell=1}^{\nb}\Var{\infden\ltrp{\rndx_{1,\ell},\rndy_{1,\ell},\hat{h}_\ell}} \label{eq:NormalVs_proof_step1} \\ 
    &=& \ns \sum_{\ell=1}^{\nb} V_s\ltrp{h_\ell,\hat{h}_\ell}.
    \label{eq:NormalVs_proof}
\end{IEEEeqnarray}
Note that in~\eqref{eq:NormalIs_proof_step1} and~\eqref{eq:NormalVs_proof_step1} we set $k=1$ without loss of
generality. We now apply the Berry-Esseen central-limit theorem~[20, Ch.~XVI.5] to the tail probability in~\eqref{eq:conditional_prob} to obtain
\begin{equation}
    \epsilon_{\text{ub}}\ltrp{\vech,\hatVech} =  Q\ltrp{ \frac{n_s
    \sum\limits_{\ell=1}^{\nb}I_s(h_{\ell},\hat{h}_\ell) - \nc\nb\rate }
{\sqrt{n_s \sum\limits_{\ell=1}^{\nb}V_s\ltrp{h_{\ell}, \hat{h}_{\ell}} } }
} + \mathcal{O}\ltrp{\frac{1}{\sqrt{n_s}}}.
    \label{eq:OutageNormalApprox}
\end{equation}
Here, both $I_s(h_\ell,\hat{h}_\ell)$ and $V_s(h_\ell,\hat{h}_\ell)$ are available in closed-form as
\begin{IEEEeqnarray}{rCl}
    I_s\ltrp{h_{\ell}, \hat{h}_{\ell}} =\ltrp{1+s\rho\abs{\hat{h}_{\ell}}^2} + (\beta_\ell - \alpha_\ell)\label{eq:Is}
\end{IEEEeqnarray}
\begin{IEEEeqnarray}{rCl}
    V_s\ltrp{h_{\ell}, \hat{h}_{\ell}} =(\beta_\ell - \alpha_\ell)^2 + 2\alpha_{\ell}\beta_{\ell} (1-\nu_\ell)\label{eq:Vs}
\end{IEEEeqnarray}
where
\begin{IEEEeqnarray}{rCL}
\alpha_\ell &=&
s\ltrp{\rho\abs{h_\ell-\hat{h}_\ell}^2+\sigma^2}\label{eq:alpha-param}\\%
\beta_\ell &=&
\frac{s}{1+s\rho\abs{\hat{h}}^2}\ltrp{\rho\abs{h_\ell}^2+\sigma^2}
\label{eq:beta-param} \\ %
 \nu_\ell &=& \frac{s^2 \abs{ \rho\abs{h_\ell}^2 + \sigma^2 - h_\ell^* \hat{h}_\ell \rho   }^2}{\alpha_\ell \beta_\ell \ltrp{1+s\rho\abs{\hat{h}_\ell}^2}}.
\label{eq:nu-param}
\end{IEEEeqnarray}

Strictly speaking, according to [20, Ch.~XVI.5], for \eqref{eq:OutageNormalApprox} to hold, we need to verify that the third central moment of $\infden\ltrp{\rndx_{1,\ell};\rndy_{1,\ell},\hat{h}_{\ell}}$
exists for every $\ell \in \{1,\dots, \nb\}$. Otherwise, the error term in \eqref{eq:OutageNormalApprox} does not vanish as $\ns$ grows, and the normal approximation is not applicable.
We will show in Section~\ref{sec:sp-ns} that the third central moment of $\infden\ltrp{\rndx_{1,\ell};\rndy_{1,\ell},\hat{h}_{\ell}}$ indeed exists.
The normal approximation w.r.t. $\ns$ is finally obtained by ignoring the
$\landauO(\cdot)$ term in \eqref{eq:OutageNormalApprox} and  by averaging 
the resulting approximation over $\rvech$ and $\hat{\rvech}$.
This is typically done via Monte-Carlo simulations.

\subsubsection{Saddlepoint Approximation w.r.t. $\ns$}\label{sec:sp-ns}
Since it is based on a central-limit theorem, the normal approximation is typically accurate only in the regime in which the target rate $\rate$ is close to the mean of the information density \cite{Lancho2020}.
However, this regime may be of limited interest in URLLC, since it may correspond
to packet error probability values above the URLLC target (see, e.g.,~\cite[Fig.~1]{ostman2020}).

A more refined approximation can be obtained by using the so-called saddlepoint
method. 
It results in an error probability expansion given in
terms of a 
leading factor that decays exponentially with $\ns$ and captures the behavior of the error probability in the large-deviation regime, and a sub-exponential factor, which is obtained by applying a refined normal approximation, and which makes the resulting approximation accurate  in the short-packet regime. 
This method allows one to obtain an approximation that is accurate for a large range of target error probabilities and rates, including the ones relevant in URLLC scenarios.

%\todo{We need to reconcile our claim that the normal approximation is not accurate, with the numerical results reported later, which show that it is accurate, actually.} With the new scaling in Fig.\ref{fig:SISO2} I think it is safe to keep this claim. (For multi-cell setup, both normal approximations are not accurate either which supports our claim even further.)

We now state this approximation. Consider again the conditional probability given in~\eqref{eq:conditional_prob}. 
We fix again $k=1$, without loss of generality, and let $\kappa(\zeta)$ be the cumulant generating function (CGF) of the random variable $-\sum_{\ell=1}^{\nb} \infden(\rndx_{1,\ell}; \rndy_{1,\ell},\hat{h}_{\ell})$:
\begin{IEEEeqnarray}{rCl}
  \kappa(\zeta) &=& \log\Ex{}{e^{-\zeta\sum_{\ell=1}^{\nb} \imath_s\ltrp{\rndx_{1,\ell}; \rndy_{1,\ell},\hat{h}_{\ell}}}} \\
                &=& \sum_{\ell=1}^{\nb} \log \Ex{}{e^{-\zeta \imath_s\ltrp{\rndx_{1,\ell}; \rndy_{1,\ell},\hat{h}_{\ell}}}}. \label{eq:CGF_outage}
\end{IEEEeqnarray}
Note that $\kappa(\zeta)$ depends on $\vech$ and $\hatVech$, but we choose not to make this dependence
explicit, to keep the notation compact.
Each term in~\eqref{eq:CGF_outage} admits a closed-form expression.
Specifically, let 
\begin{equation}\label{eq:g-fun}
  g(\zeta,h_{\ell}, \hat{h}_{\ell}) =   \Ex{}{e^{-\zeta
  \infden\ltrp{\rndx_{1,\ell};\rndy_{1,\ell},\hat{h}_{\ell}}}}
\end{equation}
be the moment generating function (MGF) of the random variable
$-\infden\ltrp{\rndx_{1,\ell};\rndy_{1,\ell},\hat{h}_{\ell}}$. 
Then~\cite[Eq.~(56)]{ostman2020}
 \begin{equation}
    \label{eq:MGFforSP}
    g(\zeta,h_{\ell},\hat{h}_{\ell})=    \frac{\ltrp{1+\ltrp{\beta_{\ell}-\alpha_{\ell}}\zeta - \alpha_{\ell}\beta_{\ell}\ltrp{1-\nu_{\ell}}\zeta^2}^{-1}}{\ltrp{1+s\rho\abs{\hat{h}_{\ell}}^2}^{\zeta}}
\end{equation}
where $\alpha_{\ell}$, $\beta_{\ell}$, and $\nu_{\ell}$ were defined
in~\eqref{eq:alpha-param},~\eqref{eq:beta-param}, and~\eqref{eq:nu-param}.

Note that, by substituting~\eqref{eq:MGFforSP} into~\eqref{eq:CGF_outage},
one can obtain a closed-form expression not only for $\kappa(\zeta)$, but also
for its first and second
derivatives, which we shall denote as $\kappa'(\zeta)$ and $\kappa''(\zeta)$,
and we shall need shortly.
Specifically,
let $\kappa_\ell(\zeta) = \log g(\zeta,h_\ell,\hat{h}_\ell)$, so that
$\kappa(\zeta) =\sum_{\ell=1}^{\nb}\kappa_{\ell} (\zeta)$. We have that (see~\cite[Eqs. (16)--(18)]{ostman2020})
\begin{IEEEeqnarray}{lCl}
  \kappa_\ell(\zeta)
     &=&{}
    -\zeta\log(1+s\rho\abs{\hat{h}_{\ell}}^2)\nonumber\\
    &&{} - \log (1+(\beta_{\ell}-\alpha_{\ell})\zeta -\alpha_{\ell}\beta_{\ell}(1-\nu_{\ell})\zeta^2)\nonumber\\\label{eq:cgf}\\
    \kappa'_\ell(\zeta) &=&{}
    -\log(1+s\rho\abs{\hat{h}_{\ell}}^2) \nonumber\\
    &&{} - \frac{(\beta_{\ell}-\alpha_{\ell}) -2\alpha\beta_{\ell}(1-\nu_{\ell})\zeta}{1+(\beta_{\ell}-\alpha_{\ell})\zeta -\alpha_{\ell}\beta_{\ell}(1-\nu_{\ell})\zeta^2} \label{eq:cgf_1} \\
    \kappa''_\ell(\zeta) &=& \biggl[\frac{(\beta_{\ell}-\alpha_{\ell}) -2\alpha_{\ell}\beta_{\ell}(1-\nu_{\ell})\zeta}{1+(\beta_{\ell}-\alpha_{\ell})\zeta -\alpha_{\ell}\beta_{\ell}(1-\nu_{\ell})\zeta^2}\biggr]^2 \nonumber\\
    &&{} +  \frac{2\alpha_{\ell}\beta_{\ell}(1-\nu_{\ell})}{1+(\beta_{\ell} - \alpha_{\ell})\zeta -\alpha_{\ell}\beta_{\ell}(1-\nu_{\ell})\zeta^2}. \label{eq:cgf_2}
\end{IEEEeqnarray} 
%

% The existence of the moment-generating function implies that the all
% moments of $-\infden(\rndx_{1,\ell}; \rndy_{1,\ell},\hat{h}_{\ell})$ are finite,
% which implies the existence of the third central moment---a condition we needed
% to establish the normal approximation~\eqref{eq:OutageNormalApprox}.
% 
A saddlepoint expansion can be established provided that the third derivative of
the MGF of $-\infden(\rndx_{1,\ell};
\rndy_{1,\ell},\hat{h}_{\ell})$ exists in a neighborhood of zero for every
$\ell\in\{1,\dots,\nb\}$. Indeed, the saddlepoint expansions presented next depend on the third derivative of the MGF, albeit this term is included in the $o(\cdot)$ and $\mathcal{O}(\cdot)$ terms and, hence, does not appear explicitly in the expansions. Specifically, for every $\ell\in\{1,\dots,\nb\}$, we require that there exist two values $\underline{\zeta}_\ell < 0 <
\overline{\zeta}_\ell$ such that
\begin{equation}
    \sup_{\underline{\zeta}_{\ell} < \zeta < \overline{\zeta}_{\ell}}
\abs{\frac{d^3}{d\zeta^3} g(\zeta,h_\ell,\hat{h}_\ell) }<\infty.\label{eq:cond_MGF_ns}
\end{equation}
As shown in~\cite[Appendix~B]{ostman2020}, this condition holds in our setup with 
\begin{IEEEeqnarray}{lCl}
  \underline{\zeta}_\ell &=& -\frac{\sqrt{(\beta_\ell-\alpha_\ell)^2 + 4\alpha_\ell\beta_\ell(1-\nu_\ell)} + \alpha_\ell - \beta_\ell}{2 \alpha_\ell \beta_\ell(1-\nu_\ell)}\label{eq:RoC_values_A}\IEEEeqnarraynumspace\\
  \overline{\zeta}_\ell &=& -\frac{\sqrt{(\beta_\ell-\alpha_\ell)^2 + 4\alpha_\ell\beta_\ell(1-\nu_\ell)} - \alpha_\ell + \beta_\ell}{2 \alpha_\ell \beta_\ell(1-\nu_\ell)}\label{eq:RoC_values_B}.\IEEEeqnarraynumspace
\end{IEEEeqnarray}
This implies in particular that the third moment of
$-\infden(\rndx_{1,\ell};\rndy_{1,\ell},\hat{h}_\ell)$, which can be obtained by
evaluating the third derivative in~\eqref{eq:cond_MGF_ns} at $\zeta=0$,
exists---a condition we required to establish the normal approximation in
Section~\ref{sec:normal-ns}.

By taking $\underline{\zeta} =
\max\{\underline{\zeta}_{1},\dots,\underline{\zeta}_{\nb}\}$ and $\overline{\zeta}
= \min\{\overline{\zeta}_{1},\dots,\overline{\zeta}_{\nb}\}$, we ensure
that~\eqref{eq:cond_MGF_ns}  holds simultaneously for every
$\ell\in\{1,\dots,\nb\}$.  
The saddlepoint expansion w.r.t. $\ns$ is stated in the following theorem.
\begin{thm}\label{thm:SPin_nc}
    Assume that there exists a $\zeta \in [\underline{\zeta},\overline{\zeta}]$
    satisfying $\rate = -
  \kappa'(\zeta)\ns/(\nc\nb)$.
If $\zeta \in [0, 1]$, then
\begin{multline}
  \epsilon_{\text{ub}}(\vech,\hatVech)= \\
e^{\ns \left(\kappa(\zeta) - \zeta\kappa'(\zeta)\right)} \left[ \Psi_{\ns,\zeta}(\zeta) + \Psi_{\ns,\zeta}(1-\zeta) + o\ltrp{\frac{1}{\sqrt{\ns}}} \right] \label{eq:RCUs_OutageR1}
\end{multline}
where 
\begin{equation}
    \label{eq:SP_Psi1}
    \Psi_{b,\zeta}(u) = e^{ b \frac{u^2}{2} \kappa''(\zeta)} Q\ltrp{u\sqrt{b \kappa''(\zeta)}}.
\end{equation} 
If $\zeta > 1$, then
\begin{multline}
  \epsilon_{\text{ub}}(\vech,\hatVech)= \\
e^{\ns\left[\kappa(1)-\kappa'(\zeta)\right]} \left[\tilde{\Psi}_{\ns}(1,1) + \tilde{\Psi}_{\ns}(0,-1) + \mathcal{O}\ltrp{\frac{1}{\sqrt{\ns}}}\right]
    \label{eq:RCUs_OutageR2}
\end{multline}
where
\begin{multline}
    \tilde{\Psi}_b(a_1,a_2) = e^{b a_1 \left[-\kappa'(1) - \rate + \frac{\kappa''(1)}{2}\right]} \\
    \times Q\ltrp{a_1\sqrt{b \kappa''(1)} - a_2\frac{b(\kappa'(1)+\rate)}{\sqrt{b\kappa''(1)}}}. 
    \label{eq:SP_Psi2}
\end{multline}
If $\zeta <0$, then
\begin{multline}
  \epsilon_{\text{ub}}(\vech,\hatVech)= 1-\\
e^{ \ns\left[\kappa(\zeta) - \zeta \kappa'(\zeta)\right]}  
\left[ \Psi_{\ns,\zeta}(-\zeta) - \Psi_{\ns,\zeta}(1-\zeta) + o\ltrp{\frac{1}{\sqrt{\ns}}} \right]. \label{eq:RCUs_OutageR3}
\end{multline}
\end{thm}
\begin{IEEEproof}
    Although a direct proof of this theorem is not available in the literature,
    the desired expansions can be established following steps similar to the
    ones
    reported in~\cite[App. E]{Scarlett2014} for the case of abstract channels
    and generic mismatch decoding rules and in~\cite[App. I]{Lancho2020} for the
    case of memoryless block-fading channels and ML decoding rule.
\end{IEEEproof} 

We obtain the desired saddlepoint approximation of
$\epsilon_{\text{ub}}(\vech,\hatVech)$ in~\eqref{eq:conditional_prob} by
omitting the $o\ltrp{\cdot}$ and the $\landauO\ltrp{\cdot}$ terms
in~\eqref{eq:RCUs_OutageR1}, \eqref{eq:RCUs_OutageR2},
and~\eqref{eq:RCUs_OutageR3}.

\subsection{Asymptotic Expansion w.r.t. $\nb$}
\label{sec:nb-expansions}
We next present a different approach, which avoids the conditioning w.r.t.
$\rvech$ and $\hat{\rvech}$ and the associated, often time-consuming, Monte-Carlo step. 
The idea is to exploit directly that the random variables
$\left\{\infden\ltrp{\rvecx_{\ell};\rvecy_{\ell},\hat{\rndh}_{\ell}}\right\}$ in~\eqref{eq:RCUsBound} 
are \iid across the block index~$\ell$, and perform an asymptotic expansion of the tail probability in~\eqref{eq:RCUsBound} w.r.t. the number of blocks $\nb$.

\subsubsection{Normal Approximation w.r.t. $\nb$}\label{sec:normal-nb}
Proceeding as in Section~\ref{sec:normal-ns}, we can obtain an asymptotic
expansion---this time directly of~$\epsilon\sub{ub}$
in~\eqref{eq:RCUsBound}---by applying the Berry-Esseen central-limit theorem. Since the random variables $\{\infden(\rvecx_\ell,\rvecy_\ell,\hat{\rndh}_\ell)\}$ are \iid in $\ell$, it follows that
\begin{IEEEeqnarray}{rCl}
    \Ex{}{\sum_{\ell = 1}^{\nb} \infden\ltrp{\rvecx_\ell;\rvecy_\ell,\hat{\rndh}_\ell}} =\nb \Ex{}{\infden\ltrp{\rvecx_1;\rvecy_1,\hat{\rndh}_1}} \label{eq:expectation_is_nb} \IEEEeqnarraynumspace
    \end{IEEEeqnarray}
and
\begin{IEEEeqnarray}{rCl}
\Var{\sum_{\ell = 1}^{\nb} \infden\ltrp{\rvecx_\ell;\rvecy_\ell,\hat{\rndh}_\ell}} = \nb \Var{\infden\ltrp{\rvecx_1;\rvecy_1,\hat{\rndh}_1}}.   \IEEEeqnarraynumspace
\label{eq:var_is_nb}
\end{IEEEeqnarray}
Here, we have fixed $\ell=1$ without loss of generality. Furthermore, since $\infden(\rvecx_1,\rvecy_1,\hat{\rndh}_1) = \sum_{k=1}^{\ns} \infden(\rndx_{k,1}; \rndy_{k,1},\hat{\rndh}_1)$ and since conditioned on $(H_1,\hat{H}_1)$, the random variables $\{\infden(\rndx_{k,1};\rndy_{k,1},\hat{\rndh}_1)\}$ are \iid in $k$, we conclude that
\begin{IEEEeqnarray}{rCl}
\IEEEeqnarraymulticol{3}{l}{
\Ex{}{\sum_{\ell = 1}^{\nb} \infden\ltrp{\rvecx_\ell;\rvecy_\ell,\hat{\rndh}_\ell}}} \nonumber \\\qquad\qquad
&=& \nb\ns \Exop \bigg[  \Ex{}{\infden\ltrp{X_{1,1};Y_{1,1},\hat{H}_{1}}\Big | \rndh_1, \hat{\rndh}_1} \bigg] \IEEEeqnarraynumspace \\
&=&  \nb\ns \Ex{}{I_s(H_{1}, \hat{H}_{1})}
\end{IEEEeqnarray}
and
\begin{IEEEeqnarray}{rCl}
\IEEEeqnarraymulticol{3}{l}{
    \Var{\sum_{\ell = 1}^{\nb} \infden\ltrp{\rvecx_\ell;\rvecy_\ell,\hat{\rndh}_\ell}} } \nonumber \\  \quad
    &=& \nb\bigg(\Exop\bigg[\Varop\Big[\infden\ltrp{\rvecx_1;\rvecy_1,\hat{\rndh}_1} 
    \Big | \rndh_1, \hat{\rndh}_1 \Big] \bigg]  
    \nonumber \\ 
    &+& \Var{\Ex{}{\infden\ltrp{\rvecx_1;\rvecy_1,\hat{\rndh}_1} \ggiven \rndh_1, \hat{\rndh}_1 }}\bigg)  \\ 
    &=& \nb \bigg(\ns \Ex{}{V_s(H_{1}, \hat{H}_{1})} + n_s^2 \Var{I_s\ltrp{H_{1}, \hat{H}_{1}}} \bigg). \IEEEeqnarraynumspace
\end{IEEEeqnarray}
It then follows that
\begin{IEEEeqnarray}{lCl}
    \label{eq:ErgodicNormalApprox}
    \epsilon_{\text{ub}} &=&{} Q\ltrp{\frac{\sqrt{\nb} \ltrp{\ns
        \Ex{}{I_s\ltrp{\rndh_1,\hat{\rndh}_1}} - \nc\rate}} { \sqrt{\ns\Ex{}{
V_s\ltrp{\rndh_1,\hat{\rndh}_1}} + \ns^2 \Var{I_s\ltrp{\rndh_1,\hat{\rndh}_1}} }} } \nonumber\\&&{} + \mathcal{O}\ltrp{\frac{1}{\sqrt{\nb}}}.
\end{IEEEeqnarray}
The normal approximation is obtained by neglecting the $\landauO(\cdot)$ term.

Note that, differently from the normal approximation provided
in~\eqref{eq:OutageNormalApprox}, 
the one provided in~\eqref{eq:ErgodicNormalApprox} applies directly to
$\epsilon\sub{ub}$ and not to the
conditional probability $\epsilon\sub{ub}(\vech,\hatVech)$. 
Hence, no
Monte-Carlo averaging step is required at the end.
On the negative side, although $I_{s}(\cdot,\cdot)$ and $V_{s}(\cdot,\cdot)$ are available
in closed form (see~\eqref{eq:Is} and~\eqref{eq:Vs}), the terms
$\Ex{}{I_s(\rndh_1,\hat{\rndh}_1)}$, $\Ex{}{ V_s(\rndh_1,\hat{\rndh}_1)}$, and
$\Var{I_s(\rndh_1,\hat{\rndh}_1)}$ need to be evaluated numerically using, e.g., Monte-Carlo methods.
Similarly, the existence of the third central moment of
$\infden\ltrp{\rvecx_{1};\rvecy_{1},\hat{\rndh}_{1}}$, which is
required for~\eqref{eq:ErgodicNormalApprox} to hold, needs to be ensured with numerical methods.

\subsubsection{Saddlepoint Approximation w.r.t. $\nb$}\label{sec:sp-nb}
We now proceed as in Section~\ref{sec:sp-ns} and obtain a saddlepoint
approximation of $\epsilon\sub{ub}$ w.r.t. $\nb$.
As pointed out in Section~\ref{sec:sp-ns}, to establish a saddlepoint asymptotic expansions we need that the third derivative of the MGF of the random variables at hand exists in a neighborhood of zero. 
Specifically, to establish a saddlepoint approximation of $\epsilon\sub{ub}$
w.r.t. $\nb$, we shall require that, for some $\underline{\zeta}< 0 <
\overline{\zeta}$,
\begin{equation}
    \label{eq:SP_condition_step1}
    \sup_{\underline{\zeta} < \zeta < \overline{\zeta}}
    \abs{\frac{\dvcube}{\dv\zeta^3} \Ex{}{e^{-\zeta \infden(\rvecx_1; \rvecy_1,
    \hat{\rndh}_1)}}} < \infty. 
\end{equation}
Unfortunately, differently from~\eqref{eq:cond_MGF_ns}, the moment-generating
function $\Ex{}{e^{-\zeta \infden(\rvecx_1; \rvecy_1, \hat{\rndh}_1)}}$ is not
known in closed form.
Hence, no closed-form expressions for $\underline{\zeta}$ and $\overline{\zeta}$
are available and these quantities need to be estimated with numerical methods. 
%\footnote{In practice, we evaluate the saddlepoint approximation by sampling $\rndh$ and $\hat{\rndh}$, and the $\zeta$ is picked within a region such that the condition \eqref{eq:SP_condition_step1} is satisfied for every sample.}

To state the saddlepoint approximation, we shall need the CGF $\gamma(\zeta)$ of
the random variable $-\infden(\rvecx_1;\rvecy_1,\hat{\rndh}_1)$
\begin{equation}
  \gamma(\zeta) = \log \Ex{}{ e^{-\zeta\infden(\rvecx_1;\rvecy_1,\hat{\rndh}_1)}}
\end{equation}
and its first and second derivatives $\gamma'(\zeta)$ and $\gamma''(\zeta)$.  
Note that, given $\rndh_{1}$ and $\hat{\rndh}_{1}$, the random variable
$\infden(\rvecx_1;\rvecy_1,\hat{\rndh}_1)$ can be decomposed into the sum of $\ns$
conditionally \iid terms (see~\eqref{eq:decomposition-is-vec}).
Hence, 
\begin{IEEEeqnarray}{rCl}
\nonumber
   \gamma(\zeta) &=& \log \Ex{\rndh_1,\hat{\rndh}_1}{\prod_{k=1}^{\ns} \Ex{}{
   e^{-\zeta\infden(\rndx_{k,1};\rndy_{k,1},\hat{\rndh}_{1} )} \ggiven \rndh_1, \hat{\rndh}_1} } \\ 
                &=& \log \Ex{\rndh_1,\hat{\rndh}_1}{g(\zeta,\rndh_{1}
                ,\hat{\rndh}_{1} )^{\ns}}
\end{IEEEeqnarray}
where the function $g$ was defined in~\eqref{eq:g-fun}. 
%Let now
Let $p(\zeta)$ be the MGF of $-\infden(\rvecx_1;\rvecy_1,\hat{\rndh}_1)$. Then $p(\zeta)$ and its first and second derivatives with respect to $\zeta$ are given as
\begin{IEEEeqnarray}{lCl}
    \label{eq:p_zeta}
    p(\zeta) &=& \Ex{\rndh_{1} ,\hat{\rndh}_{1}}{g(\zeta,\rndh_{1} ,\hat{\rndh}_{1})^{\ns}}\\
    p'(\zeta) &=& \ns \Ex{\rndh_{1} ,\hat{\rndh}_{1} }{ g(\zeta,\rndh_{1} ,\hat{\rndh}_{1})^{\ns-1}
    g'(\zeta,\rndh_{1} ,\hat{\rndh}_{1})} \label{eq:p_zetaPrime}\\
    p''(\zeta) &=& \ns \ensuremath{\Exop_{\rndh_{1} ,\hat{\rndh}_{1} }}\bigg[(\ns-1)g(\zeta,\rndh_{1}
    ,\hat{\rndh}_{1} )^{\ns-2} g'(\zeta,\rndh_{1} ,\hat{\rndh}_{1} )^2 \nonumber \\ 
    && {} + g(\zeta,\rndh_{1} ,\hat{\rndh}_{1}
  )^{\ns-1}g''(\zeta,\rndh_{1} ,\hat{\rndh}_{1} )\bigg] \IEEEeqnarraynumspace  \label{eq:p_zetaPrime2}
\end{IEEEeqnarray}
where $g'$, $g''$, $p'$ and $p''$ denote the first and second derivatives with respect to $\zeta$ of the functions $g$ and $p$, respectively.
We can then write $\gamma(\zeta)$ and its first and second derivatives as
\begin{IEEEeqnarray}{rCl}
\label{eq:gamma_dif0}
  \gamma(\zeta) &=&  \log p(\zeta) \\    
  \label{eq:gamma_dif1}
  \gamma'(\zeta) &=&  \frac{p'(\zeta) }{p(\zeta)} \\
  \label{eq:gamma_dif2}
  \gamma''(\zeta) &=& \frac{ p''(\zeta) p(\zeta) - p'(\zeta)^2}{p(\zeta)^2}. 
\end{IEEEeqnarray}

We are now ready to state the saddlepoint expansion w.r.t. $\nb$.
\begin{thm}\label{thm:SPin_L} 
  Assume that there exists a $\zeta \in [\underline{\zeta},\overline{\zeta}]$
  satisfying $\rate = -\gamma'(\zeta)/\nc$.
  If $\zeta \in [0,1]$ then 
  \begin{multline}
    \epsilon\sub{ub} = 
    e^{\nb \left[\gamma(\zeta) - \zeta\gamma'(\zeta)\right]}\\ 
    \times \left[ \Phi_{\nb,\zeta}(\zeta) + \Phi_{\nb,\zeta}(1-\zeta) + o\ltrp{\frac{1}{\sqrt{\nb}}}
    \right]\label{eq:SP-erg-1}
  \end{multline}
  where 
\begin{equation}
    \label{eq:SP_Phi1}
    \Phi_{b,\zeta}(u) = e^{ b \frac{u^2}{2} \gamma''(\zeta)} Q\ltrp{u\sqrt{b \gamma''(\zeta)}}.
\end{equation} 
  If $\zeta>1$, then 
  \begin{multline}
    \epsilon\sub{ub} =
    e^{\nb \left[\gamma(1)-\gamma'(\zeta)\right]} \\ 
    \times \left[\tilde{\Phi}_{\nb}(1,1) + \tilde{\Phi}_{\nb}(0,-1) +
    \mathcal{O}\ltrp{\frac{1}{\sqrt{\nb}}}\right]\label{eq:SP-erg-2}
  \end{multline}
  where
  \begin{multline}
    \tilde{\Phi}_b(a_1,a_2) = e^{b a_1 \left[-\gamma'(1) - \rate + \frac{\gamma''(1)}{2}\right]} \\
    \times Q\ltrp{a_1\sqrt{b \gamma''(1)} - a_2\frac{b(\gamma'(1)+\rate)}{\sqrt{b\gamma''(1)}}}. 
    \label{eq:SP_Phi2}
\end{multline}
Finally, if $\zeta<0$, 
  \begin{multline}
    \epsilon\sub{ub} =  
    1- e^{ \nb\left[\gamma(\zeta) - \zeta \gamma'(\zeta)\right]} \\ 
    \times \left[ \Phi_{\nb,\zeta}(-\zeta) - \Phi_{\nb,\zeta}(1-\zeta) + o\ltrp{\frac{1}{\sqrt{\nb}}}
    \right].\label{eq:SP-erg-3} 
  \end{multline}
\end{thm}
\begin{IEEEproof}
The proof follows by combining the steps in the proofs of~\cite[App.
I]{Lancho2020} and~\cite[App. E]{Scarlett2014}. Note that the saddlepoint
approximation in~\cite{Lancho2020} was developed for Rayleigh SISO block-fading
channels but under the assumption of ML decoding.
Furthermore, this expansion was provided only for the case  $\zeta \in [0, 1]$.
The saddlepoint derived in~\cite{Scarlett2014} holds only for channels whose input
and output belong to finite-cardinality alphabets, but applies to arbitrary mismatched
decoding rules and arbitrary values of $\zeta$.
Our result is obtained by carefully combining the proof techniques used in these two papers. 
\end{IEEEproof}

We obtain the desired saddlepoint approximation of~$\epsilon\sub{ub}$ in~\eqref{eq:RCUsBound} w.r.t.
$\nb$ by neglecting the $\landauO(\cdot)$ and the $\landauo(\cdot)$ terms
in~\eqref{eq:SP-erg-1},~\eqref{eq:SP-erg-2}, and~\eqref{eq:SP-erg-3}.
Note that, differently from the asymptotic expansion provided in Theorem~\ref{thm:SPin_nc}, 
the one provided in Theorem~\ref{thm:SPin_L} applies directly to $\epsilon\sub{ub}$ and not to the
conditional probability $\epsilon\sub{ub}(\vech,\hatVech)$. 
Hence, no
Monte-Carlo averaging step is required at the end.
On the negative side, the function $\gamma(\cdot)$ and its first and second derivatives are not available in
closed form and one needs to resort to numerical methods, such as Monte-Carlo averaging, to evaluate them.
Specifically, one needs to evaluate numerically, the expectation over the channel $H_{1}$  and its estimate
$\hat{H}_{1}$ appearing in the definition of $p(\zeta)$ and of its first and second derivatives
in~\eqref{eq:p_zeta},~\eqref{eq:p_zetaPrime}, and~\eqref{eq:p_zetaPrime2}. This is all one needs to evaluate numerically, in order to compute both the normal approximation and the saddlepoint approximation w.r.t. $\nb$. Indeed, according to~\eqref{eq:gamma_dif0},~\eqref{eq:gamma_dif1},~\eqref{eq:gamma_dif2}, from $p(\zeta)$ and its first and second derivatives, one obtains $\gamma(\zeta)$ and its first and second derivatives, which are the required quantities to evaluate the saddlepoint approximation in Theorem~1. Furthermore, since $\gamma'(0) = n_s \Ex{}{I_s(\rndh_1,\hat{\rndh}_1)}$ and $\gamma''(0) = \ns \Ex{}{V_s(\rndh_1,\hat{\rndh}_1)} + \ns^2 \Var{I_s(\rndh_1,\hat{\rndh}_1)}$, one can also evaluate the normal approximation in \eqref{eq:ErgodicNormalApprox} from $p(\zeta)$ and its first and second derivatives evaluated at $\zeta=0$.

%% ----------------------------------------------------------------------------
% \subsection{Numerical Experiments}\label{sec:SISO-experiments}
% We next evaluate numerically the two saddlepoint approximations and the two
% normal approximations just introduced, with the goal of shedding light on the following three questions:

\section{Massive MIMO Network}
\label{sec:MultiCell}
In this section, we consider a multiuser massive MIMO cellular network with
$\nl$ cells, each served by a BS with $\mM$ antennas. 
We assume there are $\nk$ single-antenna users in each cell and focus on uplink
transmission.
As in Section~\ref{sec:System}, we consider transmission over memoryless
block-fading channels and use $\nc$ and $\nb$ to denote the number of symbols
per block and the number of blocks spanned by each transmitted packet, respectively. 
We denote by $\rvech_{\ell,i,k}^{j}\in \mathbb{C}^{\mM}$ the channel gain vector within the
$\ell$th fading block between user $k$ in cell $i$ and the BS in cell $j$. 
We consider a spatially correlated Rayleigh fading model where $\rvech_{\ell,i,k}^{j} \sim
\jpg(\mathbf{0}_{\mM}, \matR_{i,k}^{j})$. 
The normalized trace $\beta_{i,k}^{j}
= \tr({\matR}_{i,k}^{j})/\mM$ determines the average channel gain between
user~$k$ in cell~$i$ and the BS in cell~$j$, while the eigenstructure
of ${\matR}_{i,k}^{j}$ describes its spatial channel correlation~\cite[Sec.
2.2]{EmilBook2017}. 

\subsection{Uplink pilot transmission}
The $\np$-dimensional pilot sequence transmitted by user~$k$ in cell~$j$ during
fading block $\ell$ is denoted by the vector $\vecx_{\ell,j,k}^{(p)} \in
\mathbb{C}^{\np}$.
We assume that this vector satisfies $\vecnorm{\vecx_{\ell,j,k}^{(p)}}^2 = \np \rho$. 
Furthermore, we assume that the $\nl\nk$ users employ mutually orthogonal pilot
sequences during each fading block. 
In particular, we set $\np = \nl\nk$.
During the pilot-transmission phase, the received signal $\rmatY_{\ell,j}^{(p)}
\in \mathbb{C}^{\mM \times \np}$ at the BS serving cell $j$ for fading block $\ell$ is given by 
\begin{IEEEeqnarray}{rCl}
    \nonumber
    \rmatY_{\ell,j}^{(p)} &=& \sum_{k=1}^{\nk} \rvech_{\ell,j,k}^{j}
    \ltrp{\vecx_{\ell,j,k}^{(p)}}^T \\&&+ \sum_{i=1,i\neq j}^{\nl} \sum_{k=1}^{\nk}
    \rvech_{\ell,i,k}^{j} \ltrp{\vecx_{\ell,i,k}^{(p)}}^T + \rmatW_{\ell,j}^{(p)}
\end{IEEEeqnarray}
where $\rmatW_{\ell,j}^{(p)} \in \mathbb{C}^{\mM \times \np}$ is the additive
noise with \iid $\jpg(0,\sigma^2)$ entries.

We assume that the BS knows 
$\matR_{i,k}^{j}$, and that it can compute the MMSE channel estimates~\cite[Sec. 3.2]{EmilBook2017}
\begin{equation}
    \hat{\rvech}_{\ell,i,k}^{j} = \matR_{i,k}^{j} \matQ_{\ell,i,k}^{j}
    \ltrp{\rmatY_{\ell,j}^{(p)} \ltrp{\vecx_{\ell,i,k}^{(p)}}^* }
\end{equation}
where 
\begin{equation}
    \matQ_{\ell,i,k}^{j} = \ltrp{\sum_{i'=1}^{\nl} \sum_{k'=1}^{\nk}
    \matR_{i',k'}^{j} \ltrp{\vecx_{\ell,i',k'}^{(p)}}^H\vecx_{\ell,i,k}^{(p)} +
\sigma^2 \matI_{\mM}}^{-1}.
\end{equation}
\subsection{Uplink data transmission}
To decode the signal transmitted from user $k$ in cell $j$ over the
$\ell$th fading block, which we denote by $\vecx_{\ell,j,k}\in\mathbb{C}^{\ns}$,
where $\ns=\nc-\np$, the BS
serving cell $j$ uses the combining vector $\rvecv_{\ell,j,k}\in\mathbb{C}^{\mM}$
to compute the  received vector $\rvecy_{\ell,j,k}\in\mathbb{C}^{\ns}$ as
follows: 
\begin{IEEEeqnarray}{rCl}
\nonumber
&&\rvecy_{\ell,j,k} = \ltrp{\rvecv_{\ell,j,k}^H \rvech_{\ell,j,k}^j} \vecx_{\ell,i,k} +
    \sum_{\substack{k'=1 \\ k'\neq k}}^{\nk} \ltrp{\rvecv_{\ell,j,k}^H \rvech_{\ell,j,k'}^j}
\vecx_{\ell,j,k'} \\ 
&& \quad + \sum_{i=1, i \neq j}^{\nl} \sum_{k'=1}^{\nk}
    \ltrp{\rvecv_{\ell,j,k}^H \rvech_{\ell,i,k'}^j} \vecx_{\ell,i,k'} +
    \rvecv_{\ell,j,k}^H
    \rmatW_{\ell,j}.
    \label{eq:MultiCellRvec}
\end{IEEEeqnarray}
Here, $\rmatW_{\ell,j}\in\mathbb{C}^{\mM \times \ns}$ is the additive Gaussian
noise on the $\ell$th fading block at the BS serving cell $j$ with \iid $\jpg(0,\sigma^2)$ entries.

We assume that the BS uses multicell-MMSE combiners,
i.e., 
\begin{equation}
    \label{eq:MMSE_Combiner}
    \rvecv_{\ell,j,k} = \ltrp{\sum_{i=1}^{\nl}\sum_{k'=1}^{\nk}
    \hat{\rvech}_{\ell,i,k'}^{j} \ltrp{\hat{\rvech}_{\ell,i,k'}^{j}}^H +
\matZ_{\ell,j} }^{-1} \hat{\rvech}_{\ell,j,k}^j
\end{equation}
where
\begin{equation}
\matZ_{\ell,j} = \sum_{i=1}^{\nl}\sum_{k=1}^{\nk} \rho \np \matR_{i,k}^{j}
\matQ_{\ell,i,k}^{j} \matR_{i,k}^{j}  + \frac{\sigma^2}{\rho} \matI_{\mM}. 
\end{equation}

Note that~\eqref{eq:MultiCellRvec} has the same form as~\eqref{eq:SISOdata}.
Indeed, set $\rvecy_\ell=\rvecy_{\ell,j,k}$, $\vecx_\ell = \vecx_{\ell,j,k}$,
$\rndh_\ell = \rvecv_{\ell,j,k}^{H} \rvech_{\ell,j,k}^{j}$, $\hat{\rndh}_{\ell}
= \rvecv_{\ell,j,k}^H \hat{\rvech}_{\ell,j,k}^j$, and $\rvecw_{\ell} =
\sum_{k'=1,k'\neq k}^{\nk} \rvecv_{\ell,j,k}^H \rvech_{\ell,j,k'}^{j}
\vecx_{\ell,j,k'} + \sum_{i=1, i \neq j}^{\nl} \sum_{k'=1}^{\nk}
\rvecv_{\ell,j,k}^H \rvech_{\ell,i,k'}^j \vecx_{\ell,i,k'} + \rvecv_{\ell,j,k}^H \rmatW_{\ell,j}$. 
Note also that, given $\{\rvech_{\ell,i,k'}^j, \hat{\rvech}_{\ell,i,k'}^j \}$, the entries
of the newly defined vector $\rvecw_{\ell}$ are conditionally \iid and follow a
$\jpg(0,\sigma_{\ell}^2)$ distribution, with 
\begin{IEEEeqnarray}{rCl}
\nonumber
    \sigma_{\ell}^2 &=& \sigma^2 \vecnorm{\rvecv_{\ell,j,k}}^2 + \rho
    \sum_{k'=1, k'\neq k }^{\nk} \abs{\rvecv_{\ell,j,k}^H \rvech_{\ell,j,k'}^j
}^2 \\ &&+ \rho \sum_{i=1,i\neq j}^{\nl} \sum_{k'=1}^{\nk}
\abs{\rvecv_{\ell,j,k}^H \rvech_{\ell,i,k'}^j}^2. 
\end{IEEEeqnarray}
Hence, we can evaluate the uplink per-user error probability by using the
information-theoretic bound in~\eqref{eq:RCUsBound} and its
normal and saddlepoint approximations discussed in Section~\ref{sec:non_asymp_SISO}.

% %% ----------------------------------------------------------------------------

\section{Numerical Results and Discussion}
\label{Sec:NumericalResults}
In this section, we report numerical experiments to evaluate the accuracy and the numerical complexity of the introduced normal and saddlepoint approximations. Specifically, we will address the following three questions:

\begin{enumerate}
    \item In typical scenarios, the number $\ns$ of symbols per block is much larger than the number $\nb$ of
    blocks spanned by a codeword. How large should $\nb$ be for the approximations w.r.t. to $\nb$ to be accurate?    
\item Is the normal approximation (either w.r.t. $\ns$ or w.r.t. $\nb$)
        sufficiently accurate in the URLLC regime, or should one use instead the
        saddlepoint approximations? 
    \item All the approximations presented in Section~\ref{sec:non_asymp_SISO}
        require numerical methods such as Monte-Carlo averaging for the
        evaluation of terms that are not available in closed form. 
        Which method has lower complexity for a given targeted accuracy?
\end{enumerate}
In the following sections, we perform numerical experiments on SISO and massive MIMO setups to answer these questions.
\subsection{SISO Setup}
\label{sec:SISO-experiments}
We start by considering a Rayleigh-fading scenario where the $\{\rndh_{\ell}\}_{\ell=1}^{\nb}$ are generated
independently from a $\jpg(0,1)$ distribution. % and the noise is independent of the channel.
Furthermore, we assume ML estimation of the channel at the receiver.
Specifically, we set 
\begin{equation}
    \hat{\rndh}_{\ell}  = \frac{1}{\rho \np} \ltrp{\vecx_{\ell}^{(p)}}^{H} \rvecy_{\ell}^{(p)}.
\end{equation}
We assume that the noise variance $\sigma_\ell^2$ is equal to $1$ for $\ell\in\{1,\dots,\nb\}$,
%variance $\sigma_\ell^2$ of the entries of the noise vector in~\eqref{eq:io-relation-siso} is equal to $1$, $\ell\in\{1,\dots,\nb\}$,
and consider a blocklength $\nb\nc$ of $288$ channel uses. 
The results reported in this section are obtained after an optimization over the parameter $s>0$ in~\eqref{eq:RCUsBound} and over the number of pilots $\np$ within each block of $\nc$ channel uses.
\begin{figure}[t]
    \centering
    \includegraphics[width=1\columnwidth]{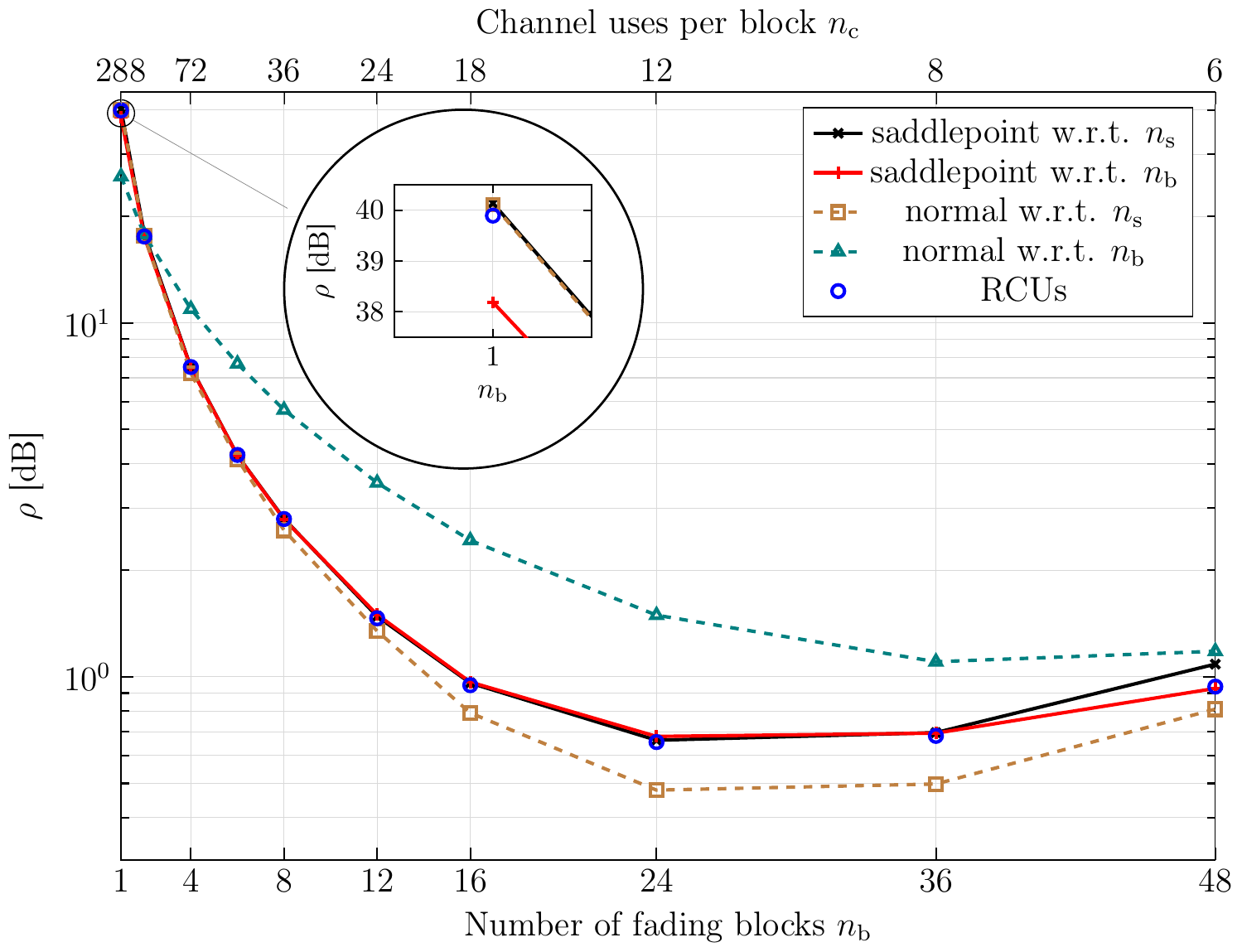}
    \caption{Upper bound on the required transmit power $\rho$ to achieve $\epsilon = 10^{-5}$.
    Here, $\nb\nc=288$, $\rate = 0.104$ bit per channel use; $n_p$ and $s$ are optimized.} 
    \label{fig:SISO2}
\end{figure}
%% ----------------------------------------------------------------------------
\paragraph*{Accuracy}
In Fig.~\ref{fig:SISO2}, we report the transmit-power value $\rho$ needed to
achieve an error probability $\epsilon=10^{-5}$ for $\rate = 0.104$ bit per
channel use, as a
function of the number of fading blocks $\nb$ spanned by each codeword.
Note that since the blocklength is fixed, $\nc$ decreases when $\nb$ is increased. 
This implies that fewer symbols are available in each block for pilot and data transmissions.
The value of $\rho$ is estimated by means of the RCUs bound
in~\eqref{eq:RCUsBound}, evaluated via a Monte-Carlo simulation
involving the generation of $2\times 10^{10}$ real Gaussian random variables,
as well as via its normal and saddlepoint approximations w.r.t. $\ns$ and w.r.t.
$\nb$, in which all expectations that need to be evaluated numerically are
computed via Monte-Carlo averaging $4\times 10^{8}$ real Gaussian random variables.

We see from the figure that the saddlepoint approximation
w.r.t. $\ns$ is accurate for $\nb$ as high as $36$, which corresponds to
$\nc=8$ symbols per block, optimally split into $\np=3$ pilots and
$\ns=5$ data symbols. 
This implies that $\ns=5$ is sufficient for the saddlepoint approximation w.r.t. $\ns$ to be accurate for this setup.
If $\nb$ is increased further, and, hence, $\nc$ is reduced, this approximation loses accuracy.
The normal approximation w.r.t. $\ns$ is accurate only for $\nb\leq 8$.   

Moving to the approximations w.r.t. to $\nb$, we note that the normal approximation does not provide accurate results
even when $\nb=48$. 
The saddlepoint approximation slightly underestimates the required transmit power $\rho$ for $\nb=1$, but, perhaps
surprisingly, returns accurate results already for $\nb$ as small as $2$. 

Note finally that for the scenario considered in the figure, the required $\rho$
is large for $\nb=1$ and decreases
rapidly until $\nb=24$, after which it increases again.
This behavior can be explained as follows.
Increasing $\nb$ for a fixed product $\nb\nc$ yields an increase of the number
of diversity branches, which is beneficial, but also of the total number of
pilot symbols $\np\nb$ which is detrimental because it increases the effective
rate of the channel code one can use to protect the information bits.
The first effect dominates for $\nb\leq 24$, whereas the second effect dominates
when $\nb>24$.  

% %
\begin{figure}[t]
    \centering
    \includegraphics[width=1\columnwidth]{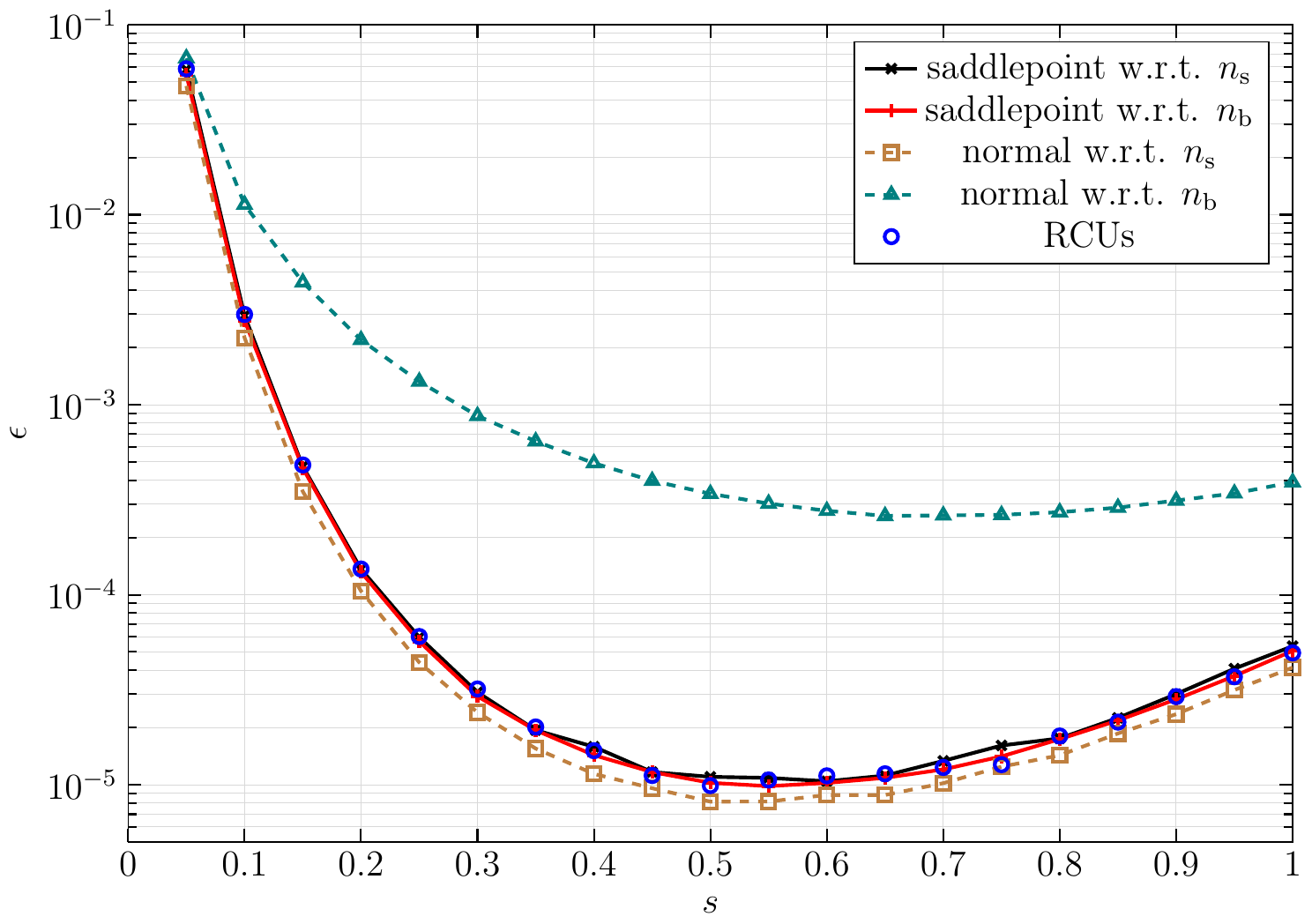}
    \caption{Packet error probability as a function of $s$. Here, $\rho=2.79 \dB$, $\nb\nc=288$, $\rate = 0.104$ bit per channel use; $\np$ is optimized. } 
    \label{fig:SISO_sbased}
\end{figure}
\begin{figure}[t]
    \centering
    \includegraphics[width=1\columnwidth]{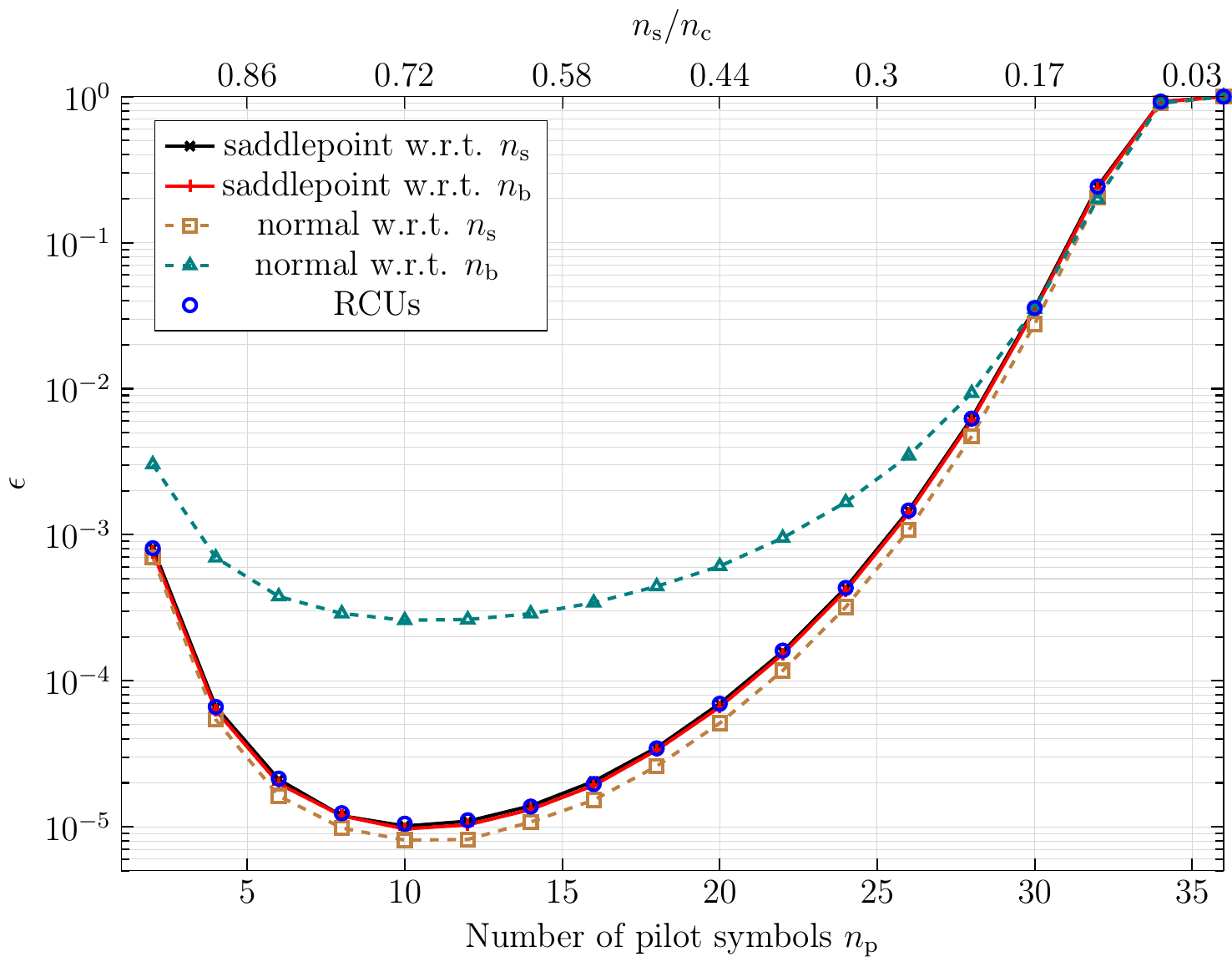}
    \caption{Packet error probability as a function of $\np$. Here, $\rho=2.79 \dB$, $\nb\nc=288$, $\rate = 0.104$ bit per channel use; $s$ is optimized. } 
    \label{fig:SISO_npBased}
\end{figure}

In Fig.~\ref{fig:SISO_sbased}, we report upper bounds on the packet-error probability $\epsilon$ as a function of $s$ for $\rate = 0.104$ bit per channel use, $\nb=8$, $\nc\nb = 288$ and $\rho = 2.79\dB$. Here, we see that all approximations of the RCUs bound with the exception of the normal approximation w.r.t. $\nb$ are accurate for a large range of values of the parameter $s$. Moreover, the parameter $s$ can be optimized to obtain a tighter bound. This figure also shows that a moderate deviation from the optimal choice of $s$ does not affect tightness significantly.

In Fig.~\ref{fig:SISO_npBased}, we report upper bounds on the packet-error probability as a function of $\np$ for $\rate = 0.104$ bit per channel use, $\nb=8$, $\nc\nb = 288$ and $\rho = 2.79\dB$. Here, in parallel to the result in Fig.~\ref{fig:SISO_sbased}, all approximations except the normal approximation w.r.t. $\nb$, provide accurate results for a large range of values of $\np$. This is crucial since $\np$ is not necessarily optimized to minimize the packet error probability in every system. Thus, being able to use the approximations, regardless of the choice of $\np$, is highly relevant. 
%% ----------------------------------------------------------------------------
\paragraph*{Complexity}
To address the third question, we assume again that all expectations that need to be
evaluated numerically in the normal and saddlepoint approximations are computed
via Monte-Carlo averaging. 
Since the channel
and its estimate are jointly Gaussian random variables, as a proxy for numerical
complexity, we count the minimum number of real Gaussian random variables that
need to be generated to guarantee that the normalized mean squared difference
between the error-probability value returned by the considered approximation
and the actual error probability bound in~\eqref{eq:RCUsBound} is less
than a given threshold. 
Specifically, we compute each approximation $\mNsim$ times for the $\rho$ value
achieving $\epsilon\sub{ub}=10^{-5}$ in~\eqref{eq:RCUsBound},  and  let
$\epsilon^{(i)}\sub{app}(\mN)$ be the
error-probability estimate obtained in the $i$th trial, when evaluating the considered approximation
for the case in which the Monte-Carlo averaging is performed using $\mN$ real
Gaussian random variables. 
The normalized mean-squared difference is evaluated as follows:
\begin{equation}
    \label{eq:accuracyMetric}
  e(\mN) = \frac{1}{\mNsim}\sum_{i=1}^{\mNsim}\ltrp{ \frac{
      \epsilon\sub{ub} -
  \epsilon^{(i)}\sub{app}(\mN) }{\epsilon\sub{ub} }}^2. 
\end{equation}
Clearly the smaller $e(\mN)$, the higher the accuracy.

\begin{figure}[t]
    \centering
    \includegraphics[width=1\columnwidth]{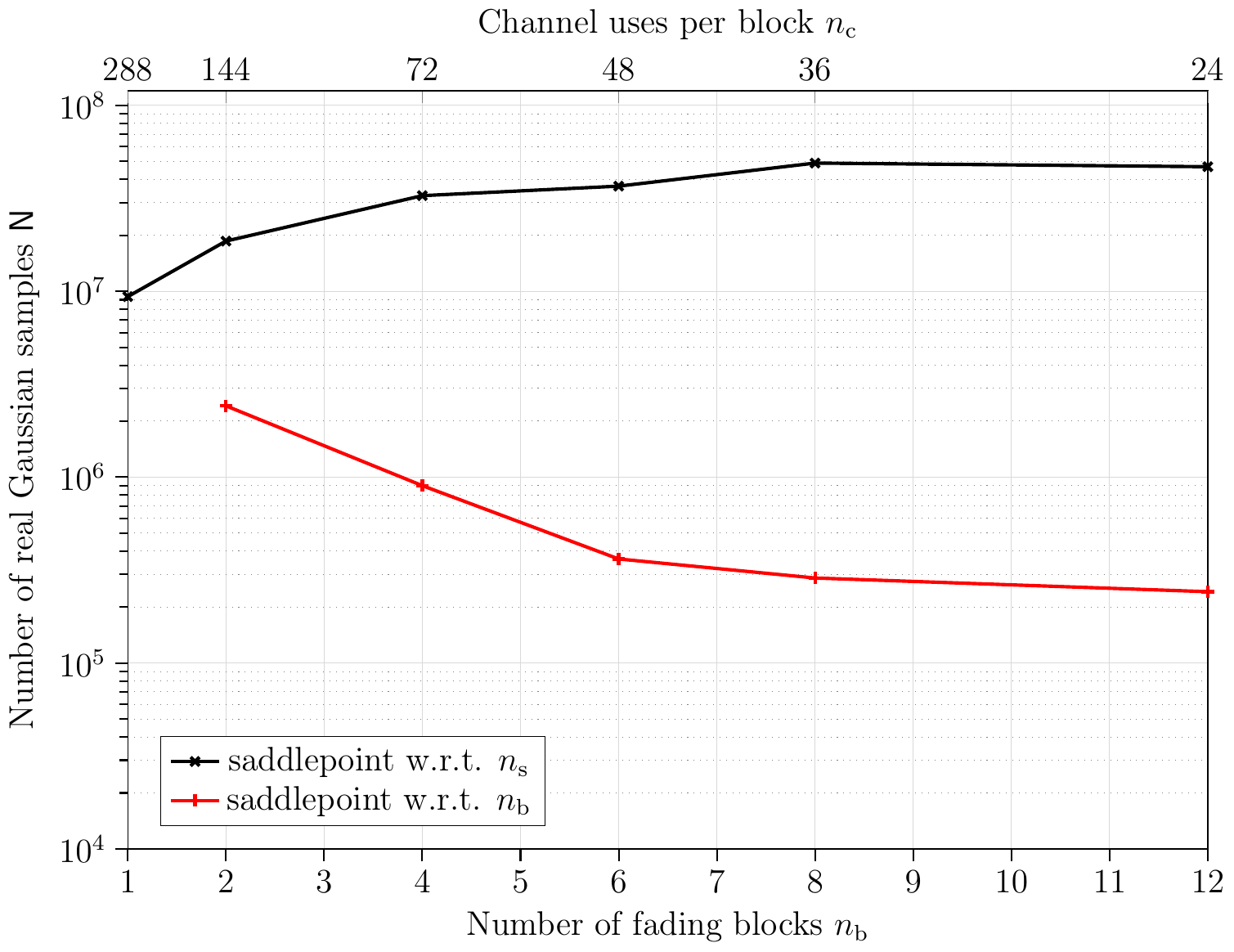}
    \caption{Required number of real Gaussian samples to guarantee that
      $e(\mN) \leq 0.5\%$. Here $\nb\nc=288$, $\rate = 0.104$ bit per channel use, $\mNsim=100$, and $\epsilon = 10^{-5}$.}
    \label{fig:SISO3}
\end{figure}

In Fig.~\ref{fig:SISO3}, we report the smallest value of $\mN$ necessary to guarantee that $
e(\mN)\leq 0.5\%$ when $\mNsim=100$, 
as a function of $\nb$.
In the figure, we assumed that $\nb\nc=288$, $\rate=0.104$ bit per channel use, and that a target error probability of $10^{-5}$
needs to be guaranteed for all values of $\nb$. 
The transmit power is set according to the RCUs curve in Fig. 1.
%This is done by adjusting the transmit power, as illustrated in Fig.~\ref{fig:SISO2}.
%
As shown in Fig.~\ref{fig:SISO3}, the saddlepoint approximations
w.r.t. $\ns$ requires around  $2 \times 10^{7}$ real Gaussian samples.
This is not surprising, since we want to evaluate accurately, via a Monte-Carlo procedure, an error
probability  of $10^{-5}$.
Although the number of random variables that need to be generated increases with $\nb$, this increase translates in a larger value of $\mN$ only for small values of $\nb$. 

On the contrary, the saddlepoint w.r.t. $\nb$ requires only around $10^{5}$ samples whenever $\nb\geq 4$.
This is around two orders of magnitude fewer samples than the saddlepoint w.r.t.~$\ns$.
This suggests that the complexity of the Monte-Carlo procedure required to evaluate numerically
the expectations in~\eqref{eq:p_zeta},~\eqref{eq:p_zetaPrime},
and~\eqref{eq:p_zetaPrime2}, 
to the level of accuracy considered in this experiment, is
much smaller than the complexity of the Monte-Carlo procedure required to evaluate numerically the
expectation over $\rvech$ and $\hat{\rvech}$ in~\eqref{eq:RCUs_ns_expect}.\footnote{Numerically achieving $e(\mN) \leq 0.5\%$ is an indication that our approximations are numerically stable.}

We do not report the complexity of the normal approximations since they do not achieve the targeted $e(\mN)$ because of their limited accuracy. 

\subsection{Massive MIMO Setup}\label{sec:numerical-experiments-2-users}
% \subsection{Numerical Experiments: A 2-User Setup}
Our simulation setup consists of $\nl$ square cells, each of size
$\SI{75}{\meter} \times \SI{75}{\meter}$, containing $\nk$ users each. 
The BSs, which are equipped with a uniform linear array with $\mM$
antenna elements separated by half a wavelength, are placed in the center of each cell. 
The antennas and the users are located in the same horizontal plane. 
Thus, the azimuth angle is sufficient to determine the directivity. 
We assume that the scatterers are uniformly distributed in the angular interval $[\varphi_{i,k} - \Delta, \varphi_{i,k} + \Delta]$,
where $\varphi_{i,k}$ is the nominal
angle of arrival of user $k$ in cell $i$ and $\Delta$ is the angular spread,
which we set to $\Delta = 25^{\circ}$. 
The $(m_{1},m_{2})$th entry of the matrix 
$\matR_{i,k}^{j}$ is then given by~\cite[Sec. 2.6]{EmilBook2017}
\begin{equation}
    \ltrsqr{\matR_{i,k}^{j}}_{m_{1},m_{2}} = \frac{\beta_{i,k}^{j}}{2\Delta}
    \int_{-\Delta}^{\Delta} e^{\mathrm{j}\pi(m_1 - m_2) \sin(\varphi_{i,k} +
    \bar{\varphi}) } \mathrm{d}\bar{\varphi}.
\end{equation}
Here, $\beta_{i,k}^{j}$ denotes the large-scale fading coefficient measured in
$\dB$
\begin{equation}
    \beta_{i,k}^{j} = -35.3 -37.6 \log_{10}\ltrp{\frac{d_{i,k}^{j}}{\SI{1}{\meter}}}
\end{equation}
with $d_{i,k}^j$ being the distance between the BS in the cell $j$ and
the user $k$ in cell $i$. 
The communication takes place over a $\SI{20}{\mega \hertz}$ bandwidth with a total receiver noise power of $\sigma^2 = -94 \dBm$ consisting of thermal noise and a noise figure of $7 \dB$ in the receiver hardware. 

In the next two subsections, we extend the accuracy and complexity analysis of the error-probability
approximations performed for a SISO link in
Section~\ref{sec:SISO-experiments} to the massive MIMO uplink. 
Since repeating the study carried out for SISO is unfeasible in a multi-cell
multi-user MIMO setting, because of complexity constraints,
we first focus in Section~\ref{sec:two-user} on a single-cell massive MIMO network with two users.
We will then provide in Section~\ref{sec:multi-user-multi-cell} an extension of this analysis to the multi-cell multiuser
massive MIMO network for the special case in which the number of blocks $\nb$ is
equal to $3$.

%% ----------------------------------------------------------------------------
\subsubsection{Accuracy and Complexity Analysis for the Two-User
Case}\label{sec:two-user}

We consider the uplink of a single-cell massive MIMO network in which the
BS serves two users ($\nl = 1$ and $\nk=2$).
The distance between the two users and the BS is $d_{1,1}^1=d_{1,2}^1 = \SI{36.4}{\meter}$.
The nominal angle of user $1$ w.r.t. the BS is $30\degree$, and the nominal
angle of user $2$ w.r.t. the BS is $40\degree$. 
We also assume that orthogonal pilot sequences are assigned to each user, that $\np=2$, and
that MMSE spatial combining based on MMSE channel estimation is used at the BS. 
Finally, we set $\nb\nc=144$ and $\rate = 2$ bit per channel use, which corresponds to 288 bits per packet.
\begin{figure}[t]
    \centering
    \includegraphics[width=1\columnwidth,keepaspectratio]{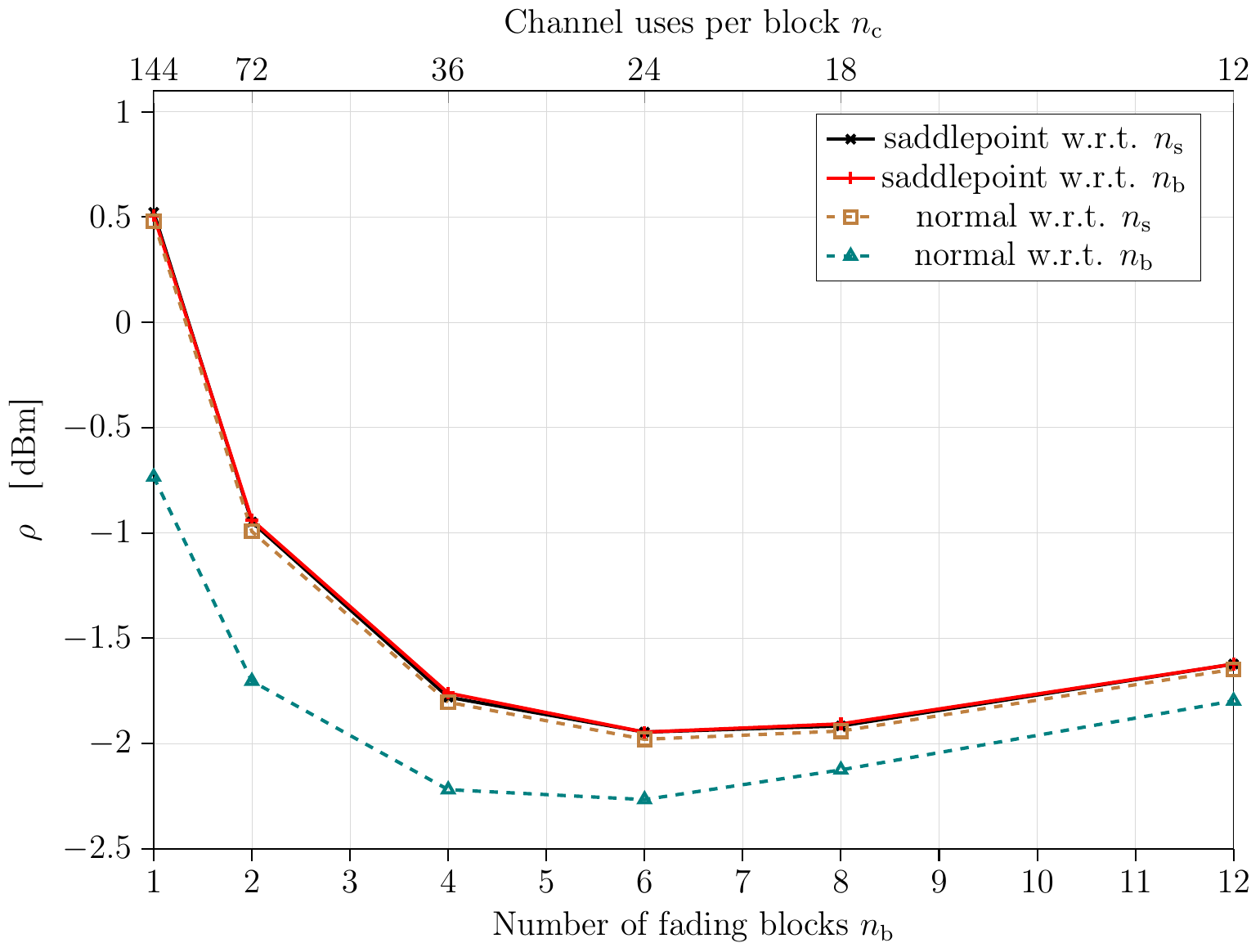}
    \caption{Two-user, single-cell massive MIMO scenario: required transmit power $\rho$ to achieve $\epsilon = 10^{-5}$. Here, $\nb\nc=144$, $\rate=2$ bit per channel use, and $\np=2$. }
    \label{fig:MIMO1}
\end{figure}
In Fig.~\ref{fig:MIMO1} we report the smallest $\rho$ value
needed to achieve an error probability of $10^{-5}$, as a function of $\nb$. 
All curves in the figures are obtained by performing a Monte-Carlo simulation
involving the generation of $8\times 10^{10}$ real Gaussian random variables.
We observe that both saddlepoint approximations as
well as the normal approximation over $\ns$ agree and are therefore assumed to be accurate for the $\nb$ values
considered in the figure, including $\nb=1$.
On the contrary, the normal approximation w.r.t. $\nb$ does not appear to be accurate.
Note that, because of the large spatial diversity available in this setup, increasing $\nb$ from $2$ to $6$ has
only a limited benefit in terms of $\rho$, and increasing $\nb$ beyond $6$ is
actually deleterious, because of the reduction in the number of channel uses per
block available
for data transmission.

\begin{figure}[t]
    \centering
    \includegraphics[width=1\columnwidth,keepaspectratio]{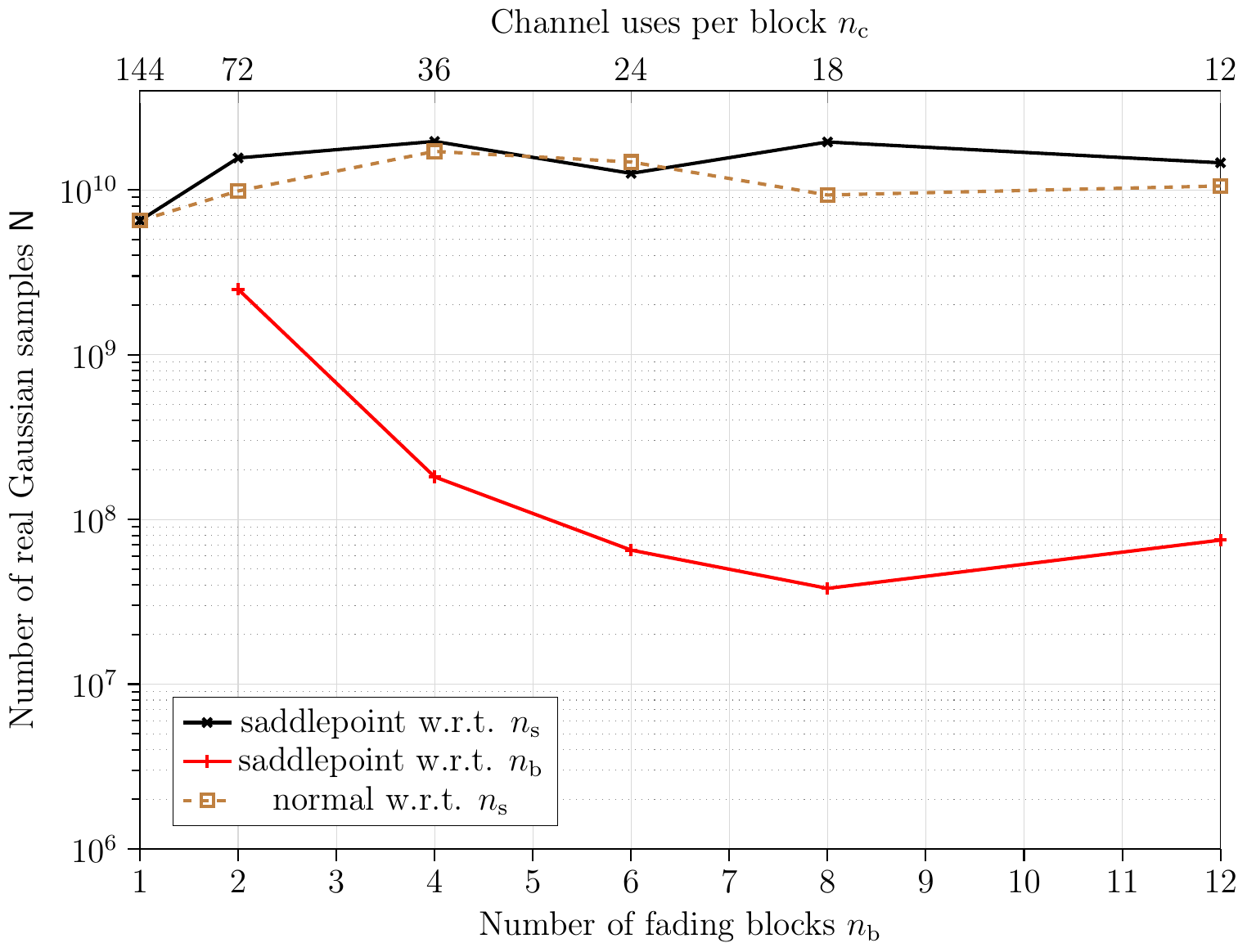}
    \caption{Two-user, single-cell massive MIMO scenario: required number of real
        Gaussian samples to guarantee that  $e(\mN) \leq
  0.5\%$. Here, $\nb\nc=144$, $\rate=2$ bit per channel use, $\np=2$, $\mNsim=100$, and $\epsilon = 10^{-5}$.}
    \label{fig:MIMO2}
\end{figure}

Focusing on both saddlepoint approximations and on the normal approximation w.r.t. $\ns$, we illustrate in Fig.~\ref{fig:MIMO2}, the minimum number of real Gaussian
samples that need to be generated in the Monte-Carlo step required in all
approximations, to guarantee that $e(\mN)< 0.5\%$ for a target error probability
of $10^{-5}$. 
Note that, unlike the SISO case, evaluating the transmit
power required to achieve $\epsilon = 10^{-5}$ with the RCUs
bound~\eqref{eq:RCUsBound} is not feasible due to its computational complexity. 
Thus, in our complexity analysis, the transmit power is set to the arithmetic average of the saddlepoint approximation curves in Fig. \ref{fig:MIMO1}.
%Thus, in our complexity analysis,  we used the arithmetic average of the transmit powers evaluated with the saddlepoint approximations w.r.t. $\ns$ and $\nb$, reported in Fig. \ref{fig:MIMO1}, as reference transmit power.

We see from Fig.~\ref{fig:MIMO2} that the number of real Gaussian samples
required by all approximations is more than two orders of magnitude larger than
in the SISO case (cf. Fig~\ref{fig:SISO3}). 
This is expected since the channel within each fading block is now characterized by $200$
(dependent) complex Gaussian random variables, instead of the single complex
Gaussian random variable needed in the SISO case. 
We also see that the saddlepoint approximation w.r.t. $\nb$ requires between $1$
and $2$ orders of magnitude fewer samples than both normal approximation and saddlepoint approximation w.r.t. $\ns$.
This observation is also in agreement with the results presented in
Section~\ref{sec:SISO-experiments} for the SISO case. 
%% ----------------------------------------------------------------------------
\subsubsection{Multi-Cell Multi-User Setup}\label{sec:multi-user-multi-cell}
We finally consider a massive MIMO network consisting of $\nl =4$ cells and $\nk
= 6$ users per cell and consider a wrap-around topology (for details, see~\cite[Sec. 4.1.3]{EmilBook2017}). 
We assume for simplicity that the users within each cell are regularly spaced
on a circle around the BS of radius $d_{j,k}^{j} = \SI{36.4}{\meter}$. 
We consider a scenario in which $\nb=3$ and $\nc=48$, assume that all users transmit orthogonal pilots over each block
and set $\np=24$. 
Finally, we set $\rate=2$ bit per channel use, and consider a target error probability of
$10^{-5}$. 
\begin{table}[t]
\renewcommand{\arraystretch}{1.3}
\caption{Required transmit powers}
\label{table:MIMO1}
\centering
\begin{tabular}{lcc}
\hline
Approximation & $\rho$\\
\hline
Saddlepoint w.r.t. $\nb$ & $-2.257\dBm$ \\
Saddlepoint w.r.t. $\ns$ & $-2.254\dBm$ \\
Normal w.r.t. $\nb$ & $-2.500\dBm$  \\
Normal w.r.t. $\ns$ & $-2.311\dBm$ \\ 
\hline
\end{tabular}
\end{table}
The required transmit powers $\rho$ for each approximation are reported in Table~\ref{table:MIMO1}.
These values of $\rho$ are estimated using a Monte-Carlo procedure involving $10^{12}$ real Gaussian random variables. 
Similar to the 2-user massive MIMO case, the RCUs bound cannot be evaluated due
to its computational complexity. 
The reported results suggest that, in agreement with the results obtained for
the SISO and for the two-user massive MIMO cases, both saddlepoint
approximations are accurate, whereas both normal approximations are not
accurate.

To assess the complexity of the two saddlepoint approximation, we take the
arithmetic average of the transmit power evaluated with the two saddlepoint
approximations as reference transmit power.
\begin{figure}[t]
    \centering
    \includegraphics[width=\columnwidth]{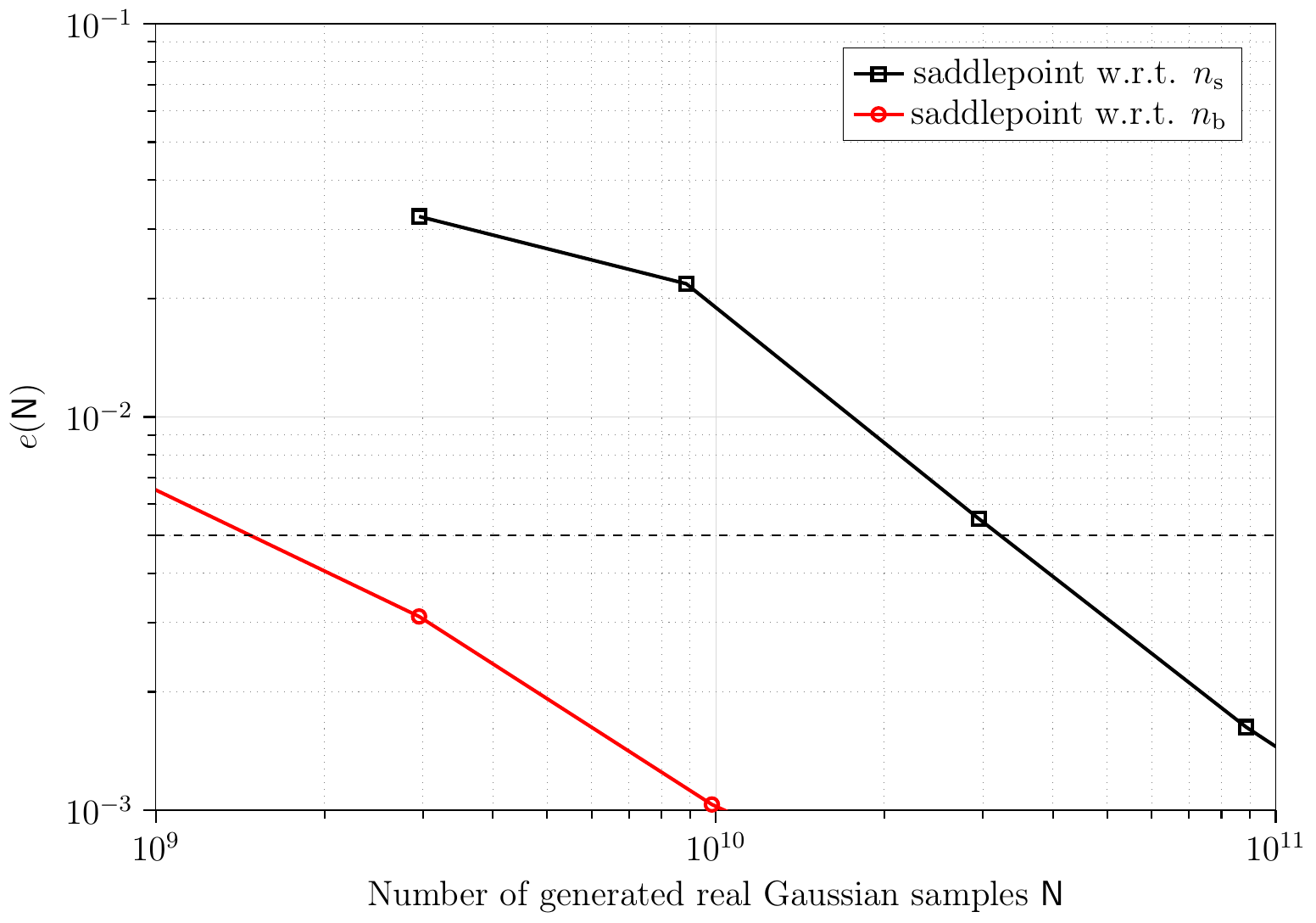}
   % \caption{Uplink transmission, $\rho \approx -1.62\dBm$, $\np=40$, $\nb\nc=210$, .}
    % \label{fig:secIIa}
  \caption{Multi-cell, multi-user massive MIMO scenario. Here, $\ns=30$,
      $\np=40$, 
      $\nb=3$, $\nk=10$, $\nl = 4$, $\mNsim=100$, $\rate = 2$ bit per channel
  use, and $\epsilon = 10^{-5}$.}\label{fig:MultiCell1}
  \end{figure}
In Fig.~\ref{fig:MultiCell1}, we depict the normalized mean-square difference
$e(\mN)$, defined in \eqref{eq:accuracyMetric}, as a function of the number of
real Gaussian samples~$\mN$ used in the Monte-Carlo step required for both
saddlepoint approximations. We can observe that the saddlepoint approximation
w.r.t. $\nb$ requires approximately $30$ times fewer samples than the
saddlepoint approximation w.r.t. $\ns$ to achieve $e(\mN) \leq 0.5\%$. This is
again in accordance with the results reported in Fig.~\ref{fig:SISO3} for the
SISO case, and Fig.~\ref{fig:MIMO2} for the single-cell, two-user massive MIMO
case. 
%% ----------------------------------------------------------------------------

\section{Conclusion}\label{sec:conclusion}
We presented numerically efficient methods to evaluate an upper bound on the error
probability achievable over SISO and massive MIMO memoryless block-fading
channels when pilot-assisted transmission, scaled nearest-neighbor decoding, and
\iid Gaussian codebooks are used. 
Our methods include both normal and saddlepoint approximations w.r.t. to the
number of data symbols per block $\ns$, as well as novel normal and saddlepoint
approximations w.r.t. the number of fading blocks $\nb$ spanned by each
codeword.
All approximations involve the numerical evaluations of expectations that are
not known in closed form and can be evaluated using Monte-Carlo methods.
Our numerical experiments reveal that the saddlepoint approximation w.r.t. to $\nb$
yield accurate estimates of the error probability in URLLC scenarios of
practical relevance.
Furthermore, it involves a 
numerical complexity (measured in terms of total number of Monte-Carlo samples
required to achieve a given accuracy) roughly two orders of magnitude lower than the complexity
of the saddlepoint approximations in $\ns$. 
This holds for a variety of scenarios ranging from SISO to multicell, multiuser
massive MIMO. 
Hence, this approximation should be preferred when evaluating error
probabilities within URLLC optimization routines such as resource-allocation and
scheduling algorithms.
The normal approximations w.r.t. $\nb$ and $\ns$ are not viable alternatives as
they often provide inaccurate results for
the scenarios considered in this paper, at no advantage in terms of complexity compared to the saddlepoint approximation. An explicit characterization of the number of samples needed for the introduced approximations to be accurate is lacking. Indeed, obtaining such a characterization would be an interesting topic for future work.

\end{document}